\documentclass[prd,twocolumn,showpacs,showkeys,superscriptaddress]{revtex4-1}   %pdftex

\usepackage{float}   % for better figure placement, [H] for forcing placement HERE
\usepackage{amsmath,amssymb}
\usepackage{graphicx}  % Include figure files
\usepackage{dcolumn}   % Align table columns on decimal point
\usepackage{bm}        % bold math
\usepackage{bbm} 
\usepackage{hyperref} % make links for the references in the document
\usepackage{color}   % for using \textcolor{color}{text}
\usepackage{pstricks}   % for using \newrgbcolor{colorname}{red level, green level, blue level}
\newcommand{\var}{\eta}

\newcommand{\standardSize}{0.45} 
\newrgbcolor{darkblue}{0 0 0.65}

\begin{document}
%\tensordelimiter{?}

\title{Circular orbits in the extreme Reissner-Nordstr{\o}m dihole metric}

%\date{today}

\author{Andreas W{\"u}nsch}
\affiliation{1. Institut f\"ur Theoretische Physik, Universit\"at Stuttgart, Pfaffenwaldring 57, 70550 Stuttgart, Germany}
\email{andreas.wuensch@itp1.uni-stuttgart.de}

\author{Thomas M{\"u}ller}
\affiliation{Visualisierungsinstitut der Universit\"at Stuttgart (VISUS), Allmandring 19, 70569 Stuttgart, Germany}
\email{thomas.mueller@vis.uni-stuttgart.de}

\author{Daniel Weiskopf}
\affiliation{Visualisierungsinstitut der Universit\"at Stuttgart (VISUS), Allmandring 19, 70569 Stuttgart, Germany}

\author{G{\"u}nter Wunner}
\affiliation{1. Institut f\"ur Theoretische Physik, Universit\"at Stuttgart, Pfaffenwaldring 57, 70550 Stuttgart, Germany}


% -----------------------------------------------------------------
%                            Abstract
% -----------------------------------------------------------------

\begin{abstract}

We study the motion of neutral test particles in the gravitational field of two charged black holes described by the extreme Reissner-Nordstr{\o}m dihole metric where the masses and charges of the black holes are chosen such that the gravitational attraction is compensated by the electrostatic repulsion. We investigate circular orbits in the equatorial plane between the two black holes with equal masses as well as the case of circular orbits outside this symmetry plane. We show that the first case reduces to an effective two-body problem with a behavior similar to a system described by the Reissner-Nordstr{\o}m spacetime. The main focus is directed to the second case with circular orbits outside the equatorial plane.
%Their existence might be counterintuitive at a first glance but can be made clear by a classical argument at least for massive particles.

\end{abstract}

\maketitle

% -----------------------------------------------------------------
%                            Introduction
% -----------------------------------------------------------------
\section{\label{sec:intro}Introduction}

The study of binary black hole systems is of great importance in the context of the emission of gravitational waves. The direct proof of the existence of such waves is still lacking and is one of the central goals in modern astrophysics. Their discovery would result in another great piece of support for the validity of Einstein's general relativity. Indirect evidence of gravitational waves has already been obtained in radio observations of binary pulsars, such as the famous binary pulsar PSR 1913+16, for which Hulse and Taylor won the Nobel Prize in physics; see~\cite{TaylorBinaryPulsar}.
%The study of binary systems of black holes is of high relevance, because of their significance for gravitational waves. A direct proof of their existence is one of the central goals in modern astrophysics and would result in another great piece of support for Einstein's general relativity. In particular, indirect evidence is possible by the study of binary systems of massive objects like neutron stars or black holes: for example, Hulse and Taylor~\cite{TaylorBinaryPulsar} analyzed the binary pulsar PSR 1913+16 and received the Nobel prize for their work in 1993. 
Furthermore, pairs of black holes could play an important part in the evolution of galaxies and the whole Universe; see for example Pallerola~\cite{Pallerola2011Diss}. 

This motivates a methodical analysis of multi black hole systems. However, for such systems no general analytic expression exists, in contrast to a single black hole. In addition, because of their character, these systems have to be described by nonstatic spacetimes. The efforts of many authors to find numerical solutions of Einstein's field equations also shows the relevance of this issue. There are, for example, approximated metrics for binary black hole systems in a circular orbit with widely separated, nonrotating black holes by Alvi~\cite{Alvi2000Approx} or a global approach by Gourgoulhon \textit{et al.}~\cite{Gourgoulhon2002Approx}. 

% One of the central goals in modern astrophysics is the direct proof of gravitational waves, what would result in another piece of great support for Einstein's general relativity. Indirect evidence is possible by the study of binary systems of massive objects like neutron stars or black holes: for example, Hulse and Taylor \cite{...} analyzed the binary pulsar PSR 1913+16 and received the Nobel prize for their work in 1993. This motivates a methodical analysis of multi black hole systems. However, for such a system no general analytic expression exists, in contrast to a single black hole. In addition, because of their character, these systems have to be described by non-static spacetimes.

The extreme Reissner-Nordstr{\o}m (RN) dihole metric describes a system of two charged black holes with compensating gravitational and electrostatic forces and is a version of the two-center problem in general relativity. Thus, a static and analytic spacetime for this special case of a double black hole system exists, which can be used as a comparatively simple model metric to analyze multi--black hole systems. 

A first step in the study of curved spacetimes is to consider the motion of test particles. The large diversity of null geodesics in the meridian plane was already pointed out by Chandrasekhar~\cite{Chandrasekhar1989}. In contrast to a corresponding Newtonian system, the relativistic motion can be chaotic as was shown by Yurtsever~\cite{Yurtsever1994Chaos}. Here, we are particularly interested in the existence of lightlike and timelike circular geodesics. Because of their simplicity, they provide a good starting point for exploring a metric. In general, they play an important role in analyzing a spacetime, as indicated by numerous publications on this topic, including Pugliese \textit{et al.}~\cite{Ruffini2011Kerr}, Chowdhury \textit{et al.}~\cite{Chowdhury2012}, or Bini \textit{et al.}~\cite{Bini2003}. 

The compensating attractive and repelling forces in the dihole system are motivated by classical mechanics. However, a strict general relativistic calculation shows the consistency of this picture with the Einstein-Maxwell equations; see~\cite{Chandrasekhar1989,ChandrasekharBuch}. Thus, in this context, it is also of interest how far the dynamics of the classical and the general relativistic systems differ. In this paper, we analyze circular orbits in the general relativistic case. A brief overview of the classical analogon for massive particles is given in Appendix~\ref{app:classicalAnalogon} and shows a remarkable relationship to its general relativistic counterpart, but also some significant differences. 

For discussing the dynamics in the extreme RN dihole spacetime, we distinguish between two cases. The first one is the motion in the equatorial plane of the extreme RN dihole metric for equal masses. In this case, we have an effective two-body problem similar to the case of a naked singularity of the simple Reissner-Nordstr{\o}m spacetime, which we will show with calculations related to the paper by Pugliese \textit{et al.}~\cite{Ruffini2011a}. They studied the existence of circular orbits in the RN metric as well as their stability for black holes and naked singularities by discussing the effective potential. They also investigated the motion of a charged particle in the RN metric~\cite{Ruffini2011b}. Our main contribution is the discussion and analysis of the second case, namely circular orbits outside the equatorial plane, which are, to the best of our knowledge, not covered in previous literature.

%% ------------------------------------------------------------------------
%%         The extreme Reissner-Nordstrom dihole metric spacetime
%% ------------------------------------------------------------------------
\section{Extreme Reissner-Nordstr{\o}m dihole metric}   \label{sec:metric}

The extreme RN dihole metric reads
\begin{equation}
   ds^2 = g_{\mu\nu}dx^{\mu}dx^{\nu} = -\frac{dt^2}{U^2}+U^2(dx^2+dy^2+dz^2)
\label{equ:MPNMetricLineElement}
\end{equation}
with
\begin{align}
   U(x,y,z) = 1 &+ \frac{M_1}{\sqrt{x^2+y^2+(z-1)^2}} \notag \\
                &+ \frac{M_2}{\sqrt{x^2+y^2+(z+1)^2}}
\label{equ:MPNMetricUFunktion}
\end{align}
(see~\cite{Chandrasekhar1989}) and describes a system of two extreme RN~black~holes with masses $M_1$ and $M_2$ at positions $\bm{r}_1=(0,0,z_1=1)$ and $\bm{r}_2=(0,0,z_2=-1)$ and charges of the same sign. The gravitational attraction is compensated by the electrostatic repulsion. For the form and properties of the RN metric, see for example~\cite{ChandrasekharBuch,Gravitation,WaldGeneralRelativity}. The metric~\eqref{equ:MPNMetricLineElement} is given in geometric units, where the speed of light $c$ and Newton's gravitational constant $G$ are normalized to unity. The extreme RN dihole metric is a special case of the Majumdar-Papapetrou spacetimes (see for example~\cite{ChandrasekharBuch}), which describe an arbitrary number of extreme RN black holes with compensating gravitational and electrostatic forces. This interpretation goes back to Hartle and Hawking~\cite{HartleHawking1972}.

Since the system~\eqref{equ:MPNMetricLineElement} is axisymmetric with respect to the $z$ axis, we use cylindrical coordinates. The metric then transforms to
\begin{equation}
   ds^2 = -\frac{dt^2}{U^2}+U^2(d\rho^2+\rho^2 d\varphi^2+dz^2)
\label{equ:MPNMetricLineElementCylindricalCoordinates}
\end{equation}
with
\begin{equation}
   U(\rho,z) = 1 + \frac{M_1}{\sqrt{\rho^2+(z-1)^2}} + \frac{M_2}{\sqrt{\rho^2+(z+1)^2}}.
\label{equ:MP2MetricUFunktionCylindricalCoordinates}
\end{equation}
We call $z=0$ the equatorial plane. A plane containing both singularities is denoted a meridian plane.

From Eq.~\eqref{equ:MPNMetricLineElementCylindricalCoordinates}, we obtain the Lagrangian of a neutral test particle
\begin{equation}
  \mathfrak{L} = \frac{1}{2}g_{\mu\nu}\dot{x}^{\mu}\dot{x}^{\nu} = \frac{1}{2} \left[ -\frac{\dot{t}^2}{U^2} + U^2(\dot{\rho}^2 + \rho^2 \dot{\varphi}^2 + \dot{z}^2) \right] .
\label{equ:LagrangeFunctionMP2MetricCylindricalCoordinates}
\end{equation}
Here, the dot denotes differentiation with respect to an affine parameter $\lambda$, which in the case of a timelike particle can be identified with its proper time. Additionally, the constraint $g_{\mu\nu}\dot{x}^{\mu}\dot{x}^{\nu} = \kappa$ has to be fulfilled, with $\kappa=0$ for null geodesics and $\kappa=-1$ for timelike geodesics. In this paper, the metric signature $\textrm{sign}(g_{\mu\nu})=+2$ is used.

Combining this constraint and Eq.~\eqref{equ:LagrangeFunctionMP2MetricCylindricalCoordinates} immediately yields $\mathfrak{L}=\kappa/2=\text{const}$. Further constants of motion follow from the cyclic coordinates $t$ and $\varphi$:
\begin{equation}
   E := \frac{\partial \mathfrak{L}}{\partial \dot{t}} = -\frac{\dot{t}}{U^2}, \quad
   L_z := \frac{\partial \mathfrak{L}}{\partial \dot{\varphi}} = \rho^2 U^2 \dot{\varphi} .
\label{equ:MP2MetricCylindricalCoordinatesConstantsOfMotion}
\end{equation}
$E$ represents the energy of a particle and $L_z$ its angular momentum with respect to the $z$ axis. With these constants, Eq.~\eqref{equ:LagrangeFunctionMP2MetricCylindricalCoordinates} can be rewritten as 
\begin{equation}
   \frac{1}{2} \dot{\rho}^2 + \frac{1}{2} \dot{z}^2 + V_{\text{eff}}(\rho,z) = \frac{E^2}{2}
\label{equ:MP2MetricCylindricalCoordinatesEnergyBalance}
\end{equation}
with the effective potential
\begin{equation}
   V_{\text{eff}}(\rho,z) = \frac{1}{2} \left( \frac{L_z^2}{\rho^2 U^4(\rho,z)} - \frac{\kappa}{U^2(\rho,z)} \right)
\label{equ:MP2MetricCylindricalCoordinatesEnergyBalanceEffectivePotential}
\end{equation}
and $U(\rho,z)$ from Eq.~\eqref{equ:MP2MetricUFunktionCylindricalCoordinates}. We choose the initial direction of a particle with respect to a local tetrad. For the metric~\eqref{equ:MPNMetricLineElementCylindricalCoordinates}, the tetrad vectors can be chosen as
\begin{eqnarray}
\begin{aligned}
  \bm{e}_{(t)} &= U \partial_t , \quad  &&\bm{e}_{(\rho)}=\frac{1}{U} \partial_{\rho} , \\
  \bm{e}_{(\varphi)} &= \frac{1}{\rho U} \partial_{\varphi} , \quad &&\bm{e}_{(z)} = \frac{1}{U} \partial_z .
\end{aligned}
\label{equ:MP2MetricCylindricalCoordinatesNaturalTetrad}
\end{eqnarray}
%
%\begin{figure}
%  \centering
%  \includegraphics[width=0.4\textwidth]{tetrad}
%  \caption{$\bm{e}_{(\rho)}$, $\bm{e}_{(\varphi)}$ and $\bm{e}_{(z)}$ define a base of the three dimensional space. The vector $\bm{n}$ can be written in spherical polar coordinates with respect to this base.}
%  \label{pic:UseOfTheNaturalTetrad}
%\end{figure}
%
Then, an initial direction $\bm{u}$ can be given by
\begin{equation}
   \bm{u}=\pm u^{(t)}\bm{e}_{(t)}+\psi\bm{n}
\label{equ:MP2MetricCylindricalCoordinatesInitialDirectionRespectivelyNaturalTetrad}
\end{equation}
with $\bm{n} = \sin\chi\cos\xi \ \bm{e}_{(\rho)}+\sin\chi\sin\xi \ \bm{e}_{(\varphi)}+\cos\chi \ \bm{e}_{(z)}$.
%\begin{equation}
%   \bm{n} = \sin\chi\cos\xi \ \bm{e}_{(\rho)}+\sin\chi\sin\xi \ \bm{e}_{(\varphi)}+\cos\chi \ \bm{e}_{(z)}.
%\label{equ:MP2MetricCylindricalCoordinatesInitialDirectionRespectivelyNaturalTetradSpacepart}
%\end{equation}
%
The sign in front of $u^{(t)}$ determines the direction of time. For a lightlike particle, we have $u^{(t)}=\psi=1$, and for a timelike particle $u^{(t)}=\gamma$ and $\psi=\beta\gamma$, with the local~velocity $\beta$ and the corresponding Lorentz factor $\gamma=(1-\beta^2)^{-1/2}$. In the case of a timelike particle, $\bm{u}$ can be identified with its four-velocity.

With respect to the initial direction~\eqref{equ:MP2MetricCylindricalCoordinatesInitialDirectionRespectivelyNaturalTetrad} the constants of motion $E$ and $L_z$ can also be written as 
\begin{equation}
   E = -\frac{\psi}{U}, \quad L_z = \psi\sin\chi \ \sin\xi \ \rho U.
\label{equ:MP2MetricCylindricalCoordinatesConstantsOfMotionRespectiveLocalTetrad}
\end{equation}

\begin{figure}
  \centering
  \includegraphics[width=\standardSize\textwidth]{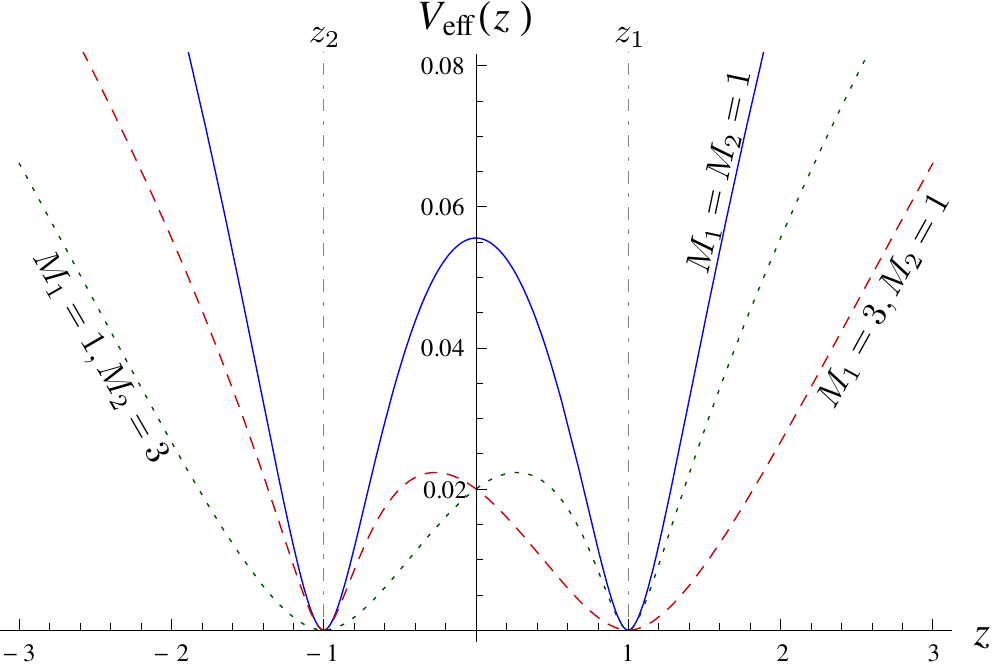}
  \caption{(color online) Effective potential $V_{\text{eff}}(z)$ from Eq.~\eqref{equ:MP2MetricEnergyBalanceAlongZAxis} for three different combinations of the mass parameters $M_1$ and $M_2$. The local maximum is always in the interval $(-1;+1)$ and marks the point of equilibrium $z_{\text{equ}}$.}
  \label{pic:effectivePotentialAlongZAxis}
\end{figure}

\begin{figure}
  \centering
  \includegraphics[width=\standardSize\textwidth]{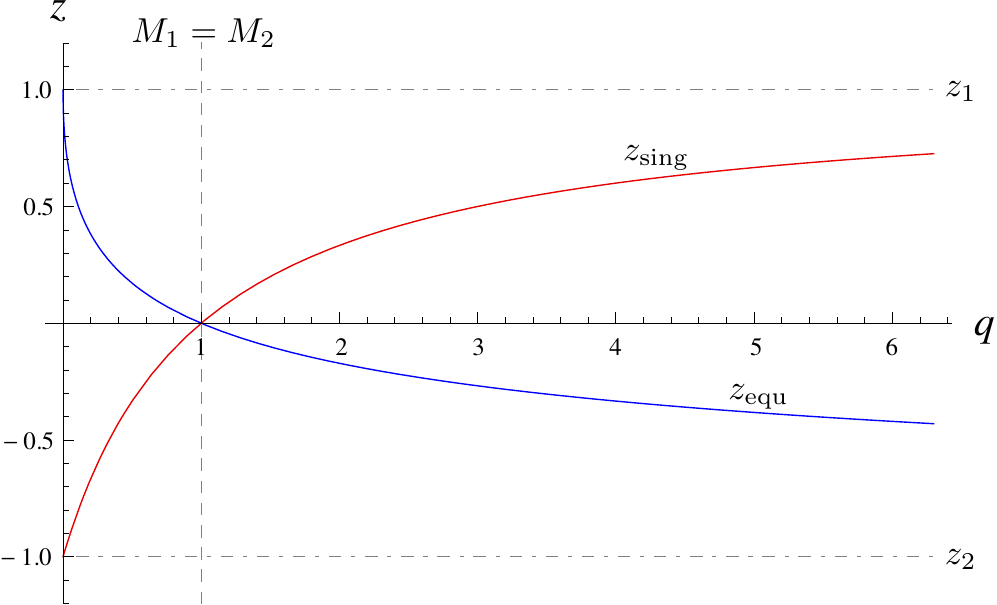}
  \caption{(color online) Point of equilibrium $z_{\text{equ}}(q)$ from Eq.~\eqref{equ:pointOfEquilibrium} and the singularity $z_{\text{sing}}(q)$ from Eq.~\eqref{equ:singularityPoint}. It is $z_{\text{equ}}(q=0)=+1$ and $z_{\text{equ}}(q\rightarrow \infty) = -1$. The bigger one mass is than the other, the closer $z_{\text{equ}}$ lies to that mass. The quantity $z_{\text{sing}}$ has exactly opposite properties and becomes important in Sec.~\ref{sec:circularOrbitsOutsideEquatorialPlane}.}
  \label{pic:pointOfEquilibrium}
\end{figure}

It is \textit{a priori} obvious that there is a point of equilibrium $z_{\text{equ}}$ on the $z$ axis, where the gravitational forces of the two masses $M_1$ and $M_2$ on a timelike test particle ($\kappa=-1$) compensate each other. A timelike particle with vanishing initial velocity at this point remains at rest, which is impossible for a lightlike particle. We can calculate $z_{\text{equ}}$ from Eq.~\eqref{equ:MP2MetricCylindricalCoordinatesEnergyBalance} by considering a test particle with initial conditions $\rho(0)=\dot{\rho}(0)=0$. In this case, $L_z$ from Eq.~\eqref{equ:MP2MetricCylindricalCoordinatesConstantsOfMotionRespectiveLocalTetrad} vanishes and thus we have
\begin{equation}
  \frac{1}{2} \dot{z}^2 + V_{\text{eff}}(z) = \frac{E^2}{2} \quad \text{with} \quad V_{\text{eff}}(z) = \frac{1}{2 U^2(z)}
\label{equ:MP2MetricEnergyBalanceAlongZAxis}
\end{equation}
and
\begin{equation}
   U(z) = 1 + \frac{M_1}{|1-z|} + \frac{M_2}{|1+z|}.
\label{equ:MP2MetricEnergyBalanceAlongZAxisEffectivePotentialUFunction}
\end{equation}
For null geodesics, the effective potential would vanish. Figure~\ref{pic:effectivePotentialAlongZAxis} shows $V_{\text{eff}}(z)$ for some combinations of the two masses $M_1$ and $M_2$.
The point of equilibrium $z_{\text{equ}}$ is given by the local maximum in the interval $(-1;+1)$. The condition $\partial V_{\text{eff}}/\partial z=0$ yields
\begin{equation}
  z_{\text{equ}}(q) = \begin{cases}  \frac{1+q - 2\sqrt{q}}{1-q} & : \quad q \neq 1 \\
                                                     0    & : \quad q = 1    \end{cases} 
\label{equ:pointOfEquilibrium}
\end{equation}
with the mass ratio $q:=M_1/M_2$. The value $z_{\text{equ}}$ depends only on the ratio of the two masses, and is shown in Fig.~\ref{pic:pointOfEquilibrium}.
%

%% ------------------------------------------------------------------------
%%         Circular orbits in the equatorial plane
%% ------------------------------------------------------------------------
\section{Circular orbits in the equatorial plane}   \label{sec:circularOrbitsInEquatorialPlane}

Since the extreme RN dihole spacetime is axisymmetric, we define a circular orbit as an axisymmetric closed orbit. Therefore, all circular geodesics fulfill $\partial V_{\text{eff}}/\partial z=0$ and $\partial V_{\text{eff}}/\partial \rho=0$ with $V_{\text{eff}}(\rho,z)$ from Eq.~\eqref{equ:MP2MetricCylindricalCoordinatesEnergyBalanceEffectivePotential}. The first condition leads to $\partial U/\partial z=0$, both for null and timelike geodesics. This results in the expression
\begin{equation}
   \frac{M_1}{M_2} = \frac{1+z}{1-z} \Bigg[1-\frac{4z}{\rho^2+(z+1)^2}\Bigg]^{3/2},
\label{equ:MP2MetricGeneralConditionForCircularOrbits}
\end{equation}
which shows that circular orbits can only exist in the range $z\in (-1;+1)$, because both masses have to be positive. Furthermore, for $z=0$, Eq.~\eqref{equ:MP2MetricGeneralConditionForCircularOrbits} becomes independent of the radius $\rho$, and it follows $M_1=M_2$. Therefore, a particle in the equatorial plane ($z=0$) with zero initial velocity components normal to this plane remains there, if the masses $M_1$ and $M_2$ are equal. In this case, we define $M:=M_1=M_2$ and Eq.~\eqref{equ:MP2MetricCylindricalCoordinatesEnergyBalance} simplifies to
\begin{equation}
   \frac{1}{2} \dot{\rho}^2 + V_{\text{eff}}(\rho) = \frac{E^2}{2} 
\label{equ:MP2MetricCylindricalCoordinatesEnergyBalanceEquatorialPlane}
\end{equation}
with
\begin{equation}
   V_{\text{eff}}(\rho) = \frac{1}{2} \left( \frac{L_z^2}{\rho^2 U^4(\rho)} - \frac{\kappa}{U^2(\rho)} \right)
\label{equ:MP2MetricCylindricalCoordinatesEnergyBalanceEquatorialPlaneEffectivePotential}
\end{equation}
and
\begin{equation}
   U(\rho) = 1 + \frac{2M}{\sqrt{\rho^2+1}}.
\label{equ:MP2MetricUFunktionCylindricalCoordinatesEquatorialPlane}
\end{equation}

\subsection{Null geodesics}   \label{sec:lightlikeCircularOrbitsInEquatorialPlane}
 
\begin{figure*}
  \centering
  \includegraphics[width=0.9\textwidth]{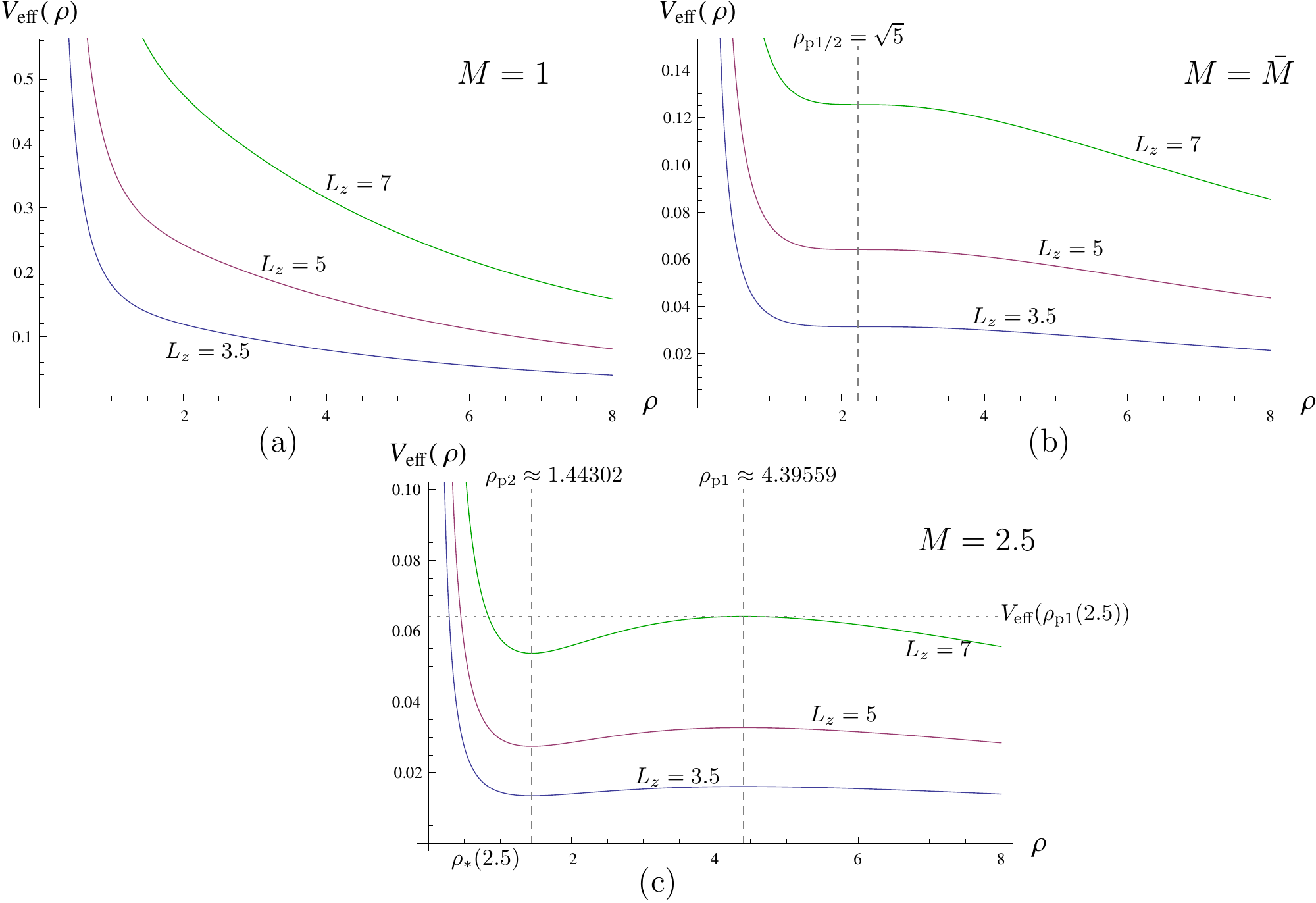}
  \caption{(color online) Effective potentials for a lightlike particle with different angular momenta $L_z$ for three different values of $M$. (a) $M=1$: There are no extremal points. (b) $M=\bar{M}\approx\sqrt{27/8}$: There is one photon orbit. (c) $M=2.5$: There are two photon orbits. The radius $\rho_*$ with $V_{\text{eff}}(\rho_*)=V_{\text{eff}}(\rho_{\text{p}1})$ and $\rho_*\neq\rho_{\text{p}1}$ is also shown.}
  \label{pic:MP2MetricPhotonOrbitsInEquatorialPlaneEffectivePotentials}
\end{figure*}
\begin{figure}[b]
  \centering
  \includegraphics[width=0.45\textwidth]{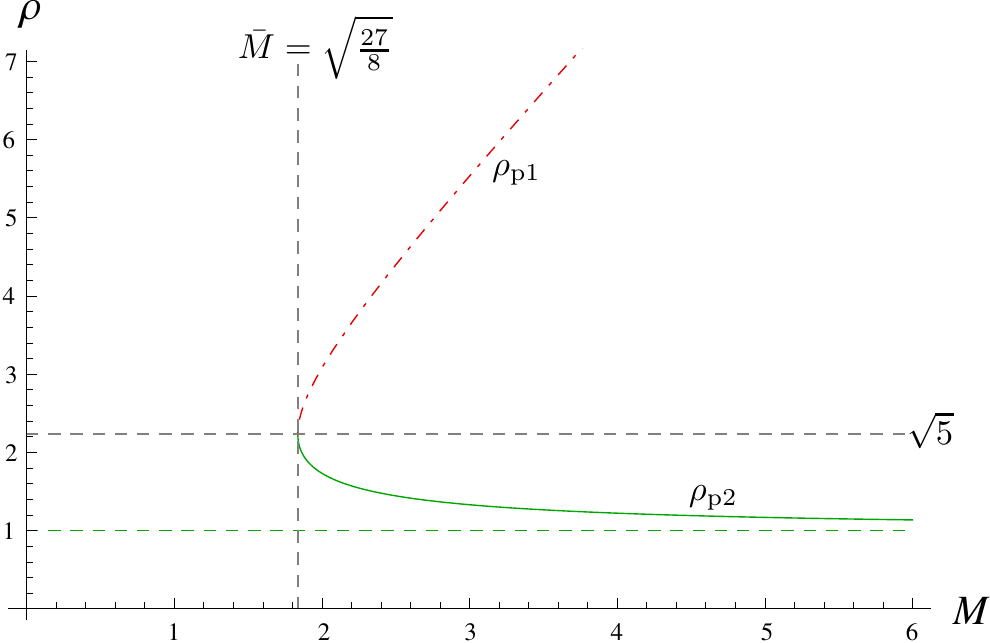}
  \caption{(color online) Photon orbits $\rho_{\text{p}1/2}(M)$ in the equatorial plane. They do not exist for $M<\bar{M}$ and coincide for $M=\bar{M}$.}
  \label{pic:MP2MetricPhotonOrbitsInEquatorialPlane}
\end{figure}

The effective potential~\eqref{equ:MP2MetricCylindricalCoordinatesEnergyBalanceEquatorialPlaneEffectivePotential} for null geodesics is shown in Fig.~\ref{pic:MP2MetricPhotonOrbitsInEquatorialPlaneEffectivePotentials} for different values of the mass $M$ and the angular momentum $L_z$. The radii of circular orbits follow from the extremal points of that effective potential. The condition $\partial V_{\text{eff}}/\partial \rho=0$ for $\kappa=0$ leads to the expression
\begin{equation}
   (\rho^2+1)^{3/2} = 2M(\rho^2-1) ,
\label{equ:MP2MetricConditionForLightlikeCircularOrbitsInEquatorialPlane}
\end{equation}
which has solutions only for $\rho\geq 1$. Substituting ${\var:=\rho^2+1}$ in Eq.~\eqref{equ:MP2MetricConditionForLightlikeCircularOrbitsInEquatorialPlane} yields the cubic equation
\begin{equation}
   \var^3 - 4M^2\var^2 + 16M^2\var - 16M^2 = 0 
\label{equ:MP2MetricConditionForLightlikeCircularOrbitsInEquatorialPlaneSubstituted}
\end{equation}
with the constraint $\var\geq 2$ because of $\rho\geq 1$. This cubic equation can be solved using Cardano's formulas. Taking $\var\geq 2$ into account, we obtain
\begin{subequations}
\begin{align}
   \var_1(M) = \frac{4}{3} &\left[ M^2 + 2M\sqrt{M^2-3} \ \cos \left( \frac{\Phi}{3} \right) \right] , \label{equ:MP2MetricLightlikeCircularOrbitsInEquatorialPlaneSubstitutedSolution1} \\
   \var_2(M) = \frac{4}{3} &\left[ M^2 - 2M\sqrt{M^2-3} \ \cos \left( \frac{\pi}{3} + \frac{\Phi}{3} \right) \right] \label{equ:MP2MetricLightlikeCircularOrbitsInEquatorialPlaneSubstitutedSolution2}
\end{align}
\label{equ:MP2MetricLightlikeCircularOrbitsInEquatorialPlaneSubstitutedSolutions}
\end{subequations}
with
\begin{equation}
   \Phi(M) := \textrm{arccos} \left( \frac{27-36M^2+8M^4}{8M\sqrt{(M^2-3)^3}} \right).
\label{equ:MP2MetricLightlikeCircularOrbitsInEquatorialPlaneSubstitutedSolutionsBigPhi}
\end{equation}
The quantities $\var_1$ and $\var_2$ are defined only for $M\geq\bar{M}$ with
\begin{equation}
   \bar{M} := \sqrt{\frac{27}{8}} \approx 1.83712,
\label{equ:MP2MetricDefinitionOfBarM}
\end{equation}
which follows from the roots of the discriminant of the cubic~\eqref{equ:MP2MetricConditionForLightlikeCircularOrbitsInEquatorialPlaneSubstituted}. The two corresponding photon orbits with radii $\rho_{\text{p}1}$ and $\rho_{\text{p}2}$ then follow from $\rho=\sqrt{\var-1}$. The radius $\rho_{\text{p}1}(M)$ is a local maximum of $V_{\text{eff}}(\rho)$ and thus the orbit is an unstable circular orbit. The radius $\rho_{\text{p}2}(M)$ is a local minimum of $V_{\text{eff}}(\rho)$ and thus the orbit is stable in a local sense along the $\rho$ direction. Both photon orbits are shown in Fig.~\ref{pic:MP2MetricPhotonOrbitsInEquatorialPlane}.

\begin{figure}[H]
  \centering
  \includegraphics[width=\standardSize\textwidth]{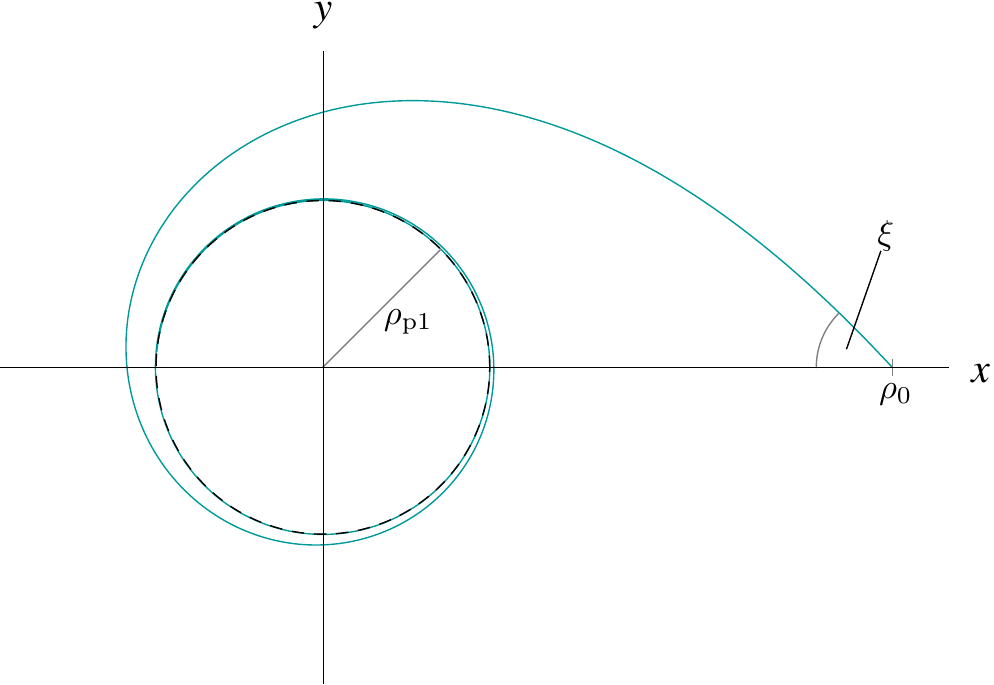}
  \caption{(color online) A light ray is emitted in the equatorial plane from $\rho_0$ at an angle $\xi$ and asymptotically approaches the photon orbit $\rho_{\text{p}1}$.}
  \label{pic:MP2MetricAsymptoticalTrajectoryToPhotonOrbitInEquatorialPlane}
\end{figure}

\begin{figure}[H]
  \centering
  \includegraphics[width=\standardSize\textwidth]{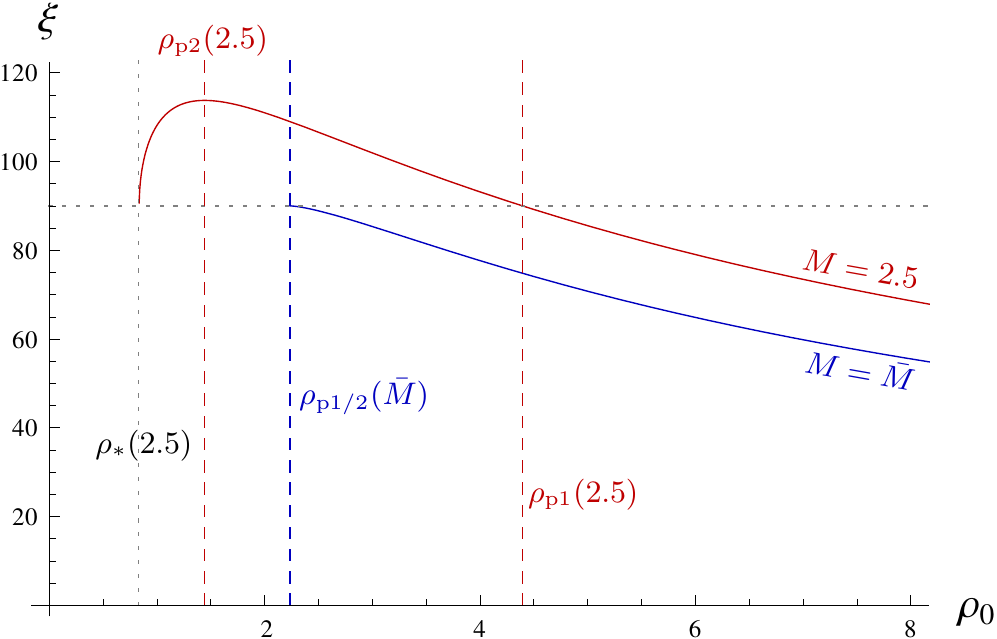}
  \caption{(color online) The angle $\xi(\rho_0)$ from Eq.~\eqref{equ:MP2MetricCriticalAngleInEquatorialPlane} in degrees for $M=\bar{M}=\sqrt{27/8}$ and $M=2.5$.}
  \label{pic:MP2MetricCriticalAngleForPhotonOrbitInEquatorialPlane}
\end{figure}

The existence of a local maximum in the effective potential implies the existence of an asymptotic trajectory to this photon orbit. To see this we consider an observer at distance $\rho_0$ in the equatorial plane that emits a light ray at angle $\xi$ (see Fig.~\ref{pic:MP2MetricAsymptoticalTrajectoryToPhotonOrbitInEquatorialPlane}). Such an asymptotic orbit has to fulfill
\begin{equation}
   V_{\text{max}} := V_{\text{eff}}(\rho_{\text{p}1}) = \frac{E^2}{2}
\label{equ:MP2MetricConditionForCriticalAngleInEquatorialPlane}
\end{equation}
[see also Fig.~\ref{pic:MP2MetricPhotonOrbitsInEquatorialPlaneEffectivePotentials}(c)].
%\begin{figure}
%  \centering
%  %\includegraphics[width=0.45\textwidth]{asymptoticalPhotonOrbit}
%  \includegraphics[width=\standardSize\textwidth]{wuensch_figure05}
%  \caption{(color online) A light ray is emitted in the equatorial plane from $\rho_0$ at an angle $\xi$ and asymptotically approaches the photon orbit $\rho_{\text{p}1}$.}
%  \label{pic:MP2MetricAsymptoticalTrajectoryToPhotonOrbitInEquatorialPlane}
%\end{figure}
%
The constants of motion $E$ and $L_z$ can be expressed with respect to the local tetrad according to Eq.~\eqref{equ:MP2MetricCylindricalCoordinatesConstantsOfMotionRespectiveLocalTetrad} with $z=0$ and $\chi=\pi/2$. Equation~\eqref{equ:MP2MetricConditionForCriticalAngleInEquatorialPlane} then reads
\begin{equation}
   \frac{\rho_0^2 U^2(\rho_0) \ \sin^2\xi}{2 \rho_{\text{p}1}^2 U^4(\rho_{\text{p}1})} = \frac{1}{2U^2(\rho_0)}
\label{equ:MP2MetricEquationForCriticalAngleInEquatorialPlane}
\end{equation}
and from this, we obtain
\begin{equation}
   \xi(\rho_0) = \begin{cases} 
       \textrm{arcsin}\left(\frac{\rho_{\text{p}1}U^2(\rho_{\text{p}1})}{\rho_0U^2(\rho_0)}\right) & : \phantom{..} \rho_0 \geq \rho_{\text{p}1} \\
   \pi-\textrm{arcsin}\left(\frac{\rho_{\text{p}1}U^2(\rho_{\text{p}1})}{\rho_0U^2(\rho_0)}\right) & : \phantom{..} \rho_0 \leq \rho_{\text{p}1} ,
                       \end{cases}
\label{equ:MP2MetricCriticalAngleInEquatorialPlane}
\end{equation}
with $U(\rho)$ from Eq.~\eqref{equ:MP2MetricUFunktionCylindricalCoordinatesEquatorialPlane}. Typical curves of $\xi(\rho_0)$ are shown in Fig.~\ref{pic:MP2MetricCriticalAngleForPhotonOrbitInEquatorialPlane}. The function is only defined if the photon orbit $\rho_{\text{p}1}$ exists. For an observer at this orbit, the angle is $\xi=90^{\circ}$ in accordance with the similar situation
\begin{figure}[H]
  \centering
  \includegraphics[width=0.48\textwidth]{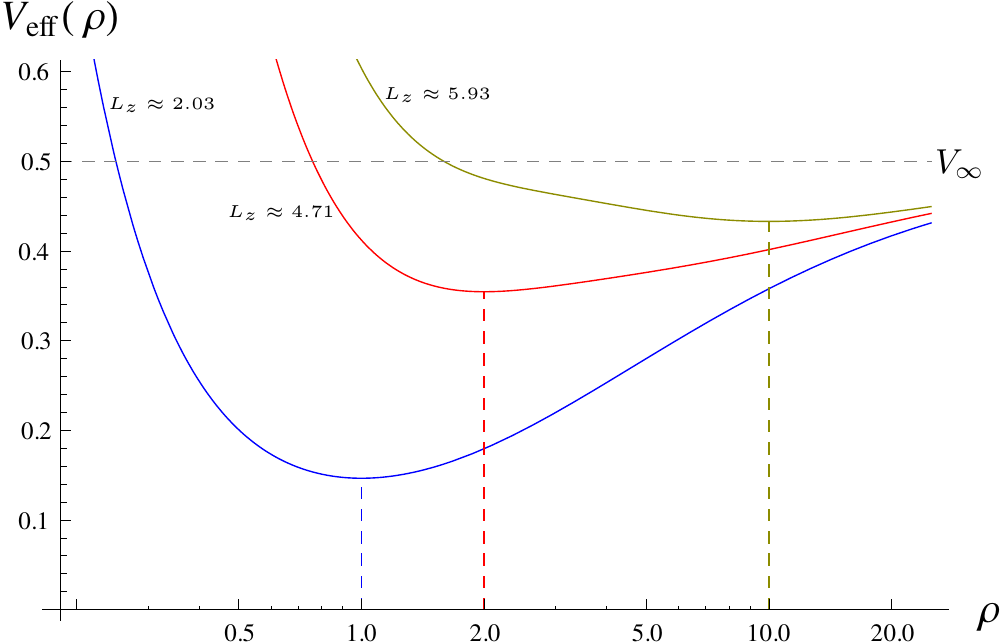}
  \caption{(color online) Effective potentials for a timelike particle in the case $M=1$ with different angular momenta $L_z$.}
  \label{pic:MP2MetricTimelikeParticleInEquatorialPlaneEffectivePotential1}
\end{figure}
\begin{figure}[H]
  \centering
  \includegraphics[width=0.455\textwidth]{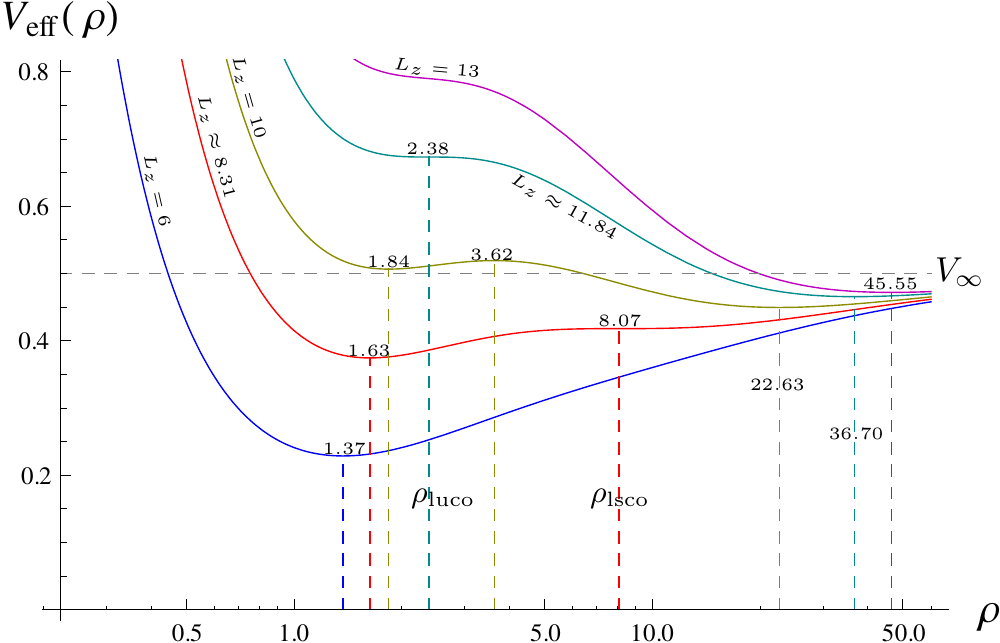}
  \caption{(color online) Effective potentials for a timelike particle in the case $M=1.5$ with different angular momenta $L_z$. In this and the following figures, the radii $\rho_{\text{lsco}}$ and $\rho_{\text{luco}}$ stand for the last radially stable and unstable circular orbits, and are obtained from Eq.~\eqref{equ:MP2MetricExtremalPointsOfMathcalLAndE}. Further information on their nomenclature is in the text.}
  \label{pic:MP2MetricTimelikeParticleInEquatorialPlaneEffectivePotential2}
\end{figure}
\begin{figure}[H]
  \centering
  \includegraphics[width=0.455\textwidth]{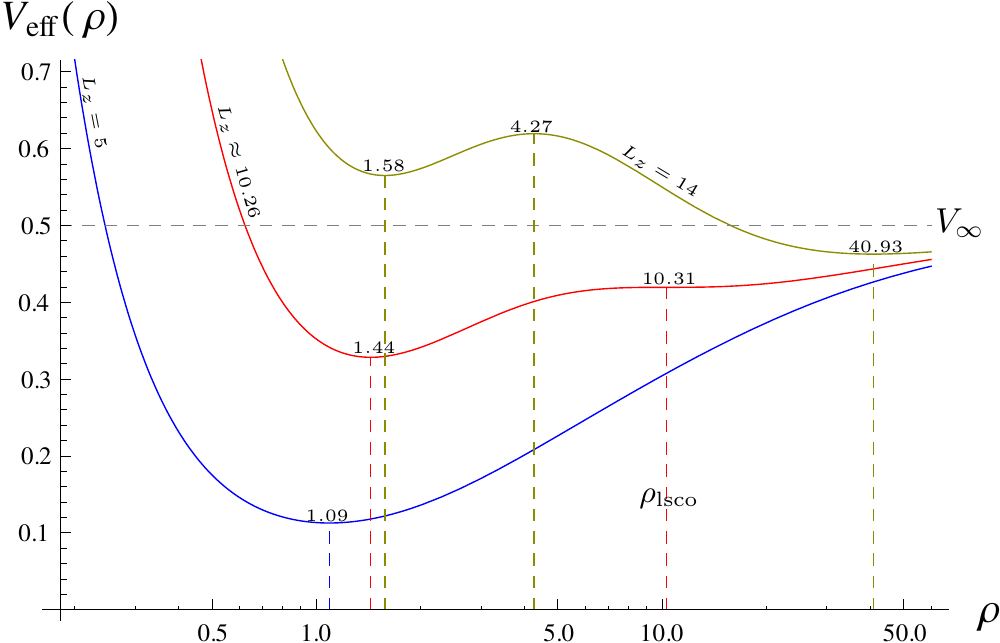}
  \caption{(color online) Effective potentials for a timelike particle in the case $M=\bar{M}=\sqrt{27/8}\approx 1.837$ with different angular momenta $L_z$.}
  \label{pic:MP2MetricTimelikeParticleInEquatorialPlaneEffectivePotential3}
\end{figure}
\begin{figure}[H]
  \centering
  \includegraphics[width=0.455\textwidth]{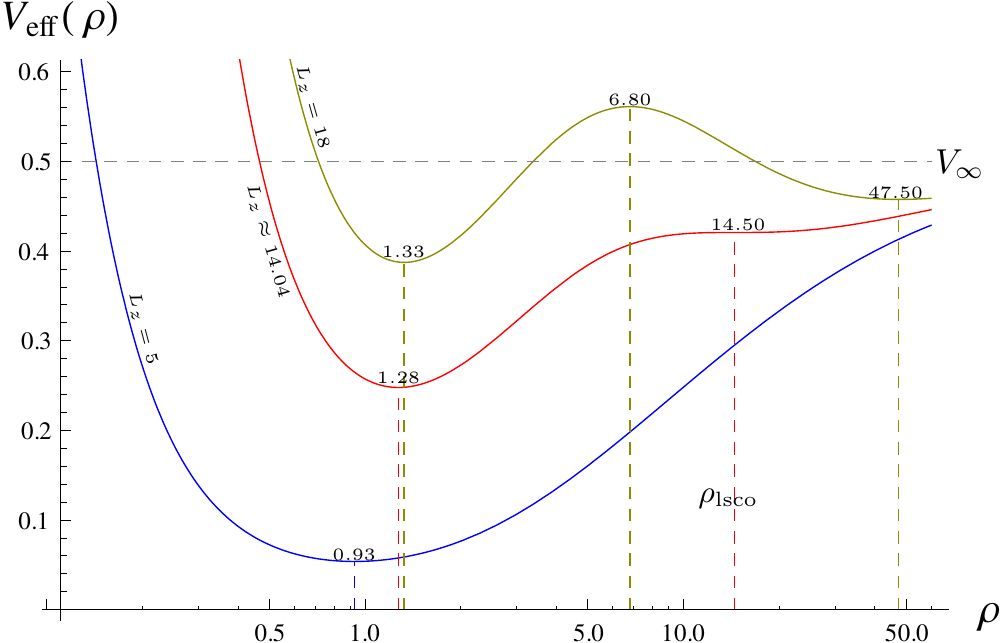}
  \caption{(color online) Effective potentials for a timelike particle in the case $M=2.5$ with different angular momenta $L_z$.}
  \label{pic:MP2MetricTimelikeParticleInEquatorialPlaneEffectivePotential4}
\end{figure}
\clearpage
\begin{figure}[H]
  \centering
  \includegraphics[width=\standardSize\textwidth]{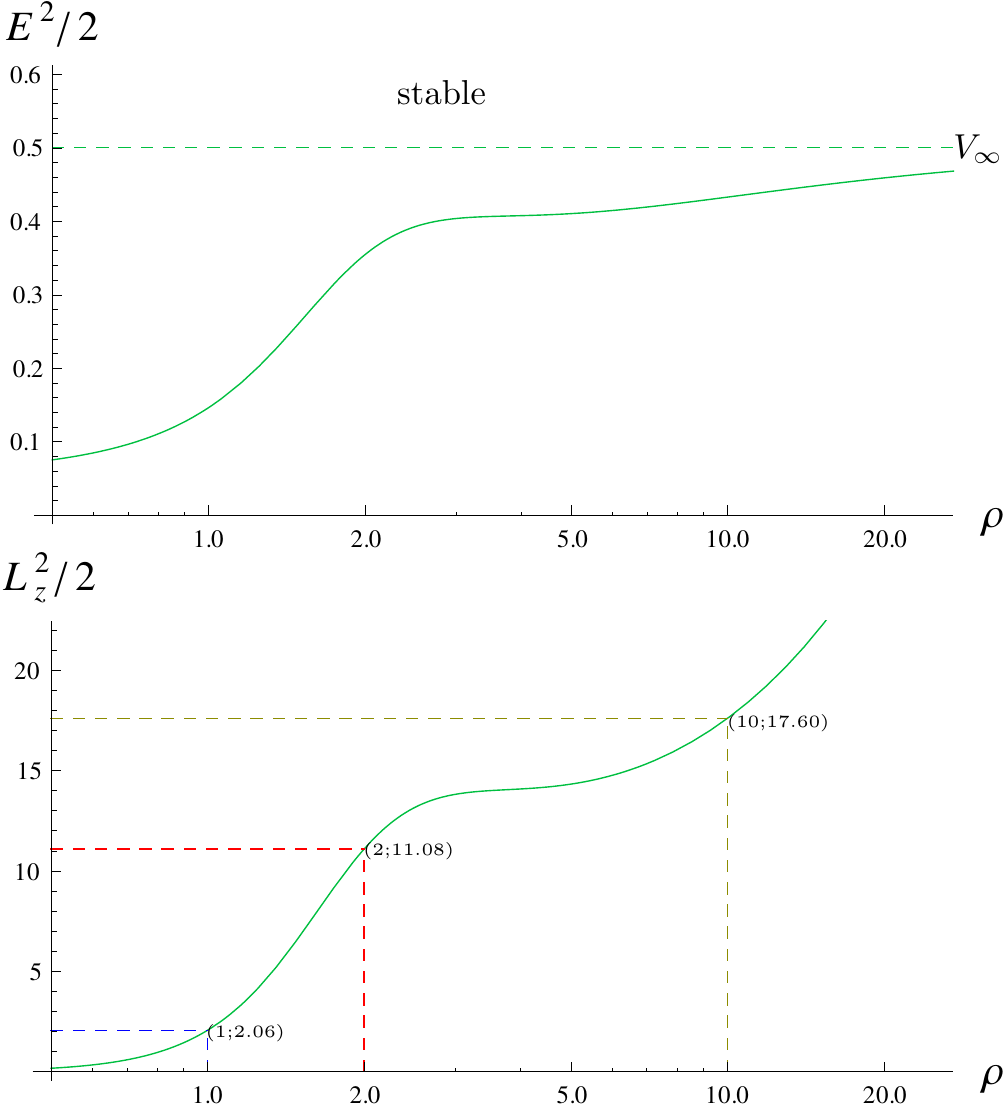}
  \caption{(color online) The functions $E^2(\rho)/2$ and $L_z^2(\rho)/2$ for $M=1$ as an example of the interval $0<M\leq M_*\approx 1.029$ [see Eq.~\eqref{equ:MP2MetricDefinitionOfMStar}]. For all $L_z^2/2$, there is one circular orbit (see also Fig.~\ref{pic:MP2MetricTimelikeParticleInEquatorialPlaneEffectivePotential1}). For $M=M_*$, there would be a reflection point in the diagrams. In this and the following figures, the stability refers to the radial direction.}
  \label{pic:MP2MetricTimelikeParticleInEquatorialPlaneEnergyAndAngularMomentum1}
\end{figure}
\noindent
in the Schwarzschild metric (compare with M{\"u}ller~\cite{MuellerBoblest2010}), which can also be easily expanded to the RN metric.

\subsection{Timelike geodesics}   \label{sec:timelikeCircularOrbitsInEquatorialPlane}

The circular orbits of a timelike test particle follow from the extremal points of the effective potential in Eq.~\eqref{equ:MP2MetricCylindricalCoordinatesEnergyBalanceEquatorialPlaneEffectivePotential} for $\kappa=-1$. This function is shown in Figs.~\ref{pic:MP2MetricTimelikeParticleInEquatorialPlaneEffectivePotential1}--\ref{pic:MP2MetricTimelikeParticleInEquatorialPlaneEffectivePotential4} for different masses $M$ and angular momenta $L_z$.

\begin{figure}
  \centering
  \includegraphics[width=\standardSize\textwidth]{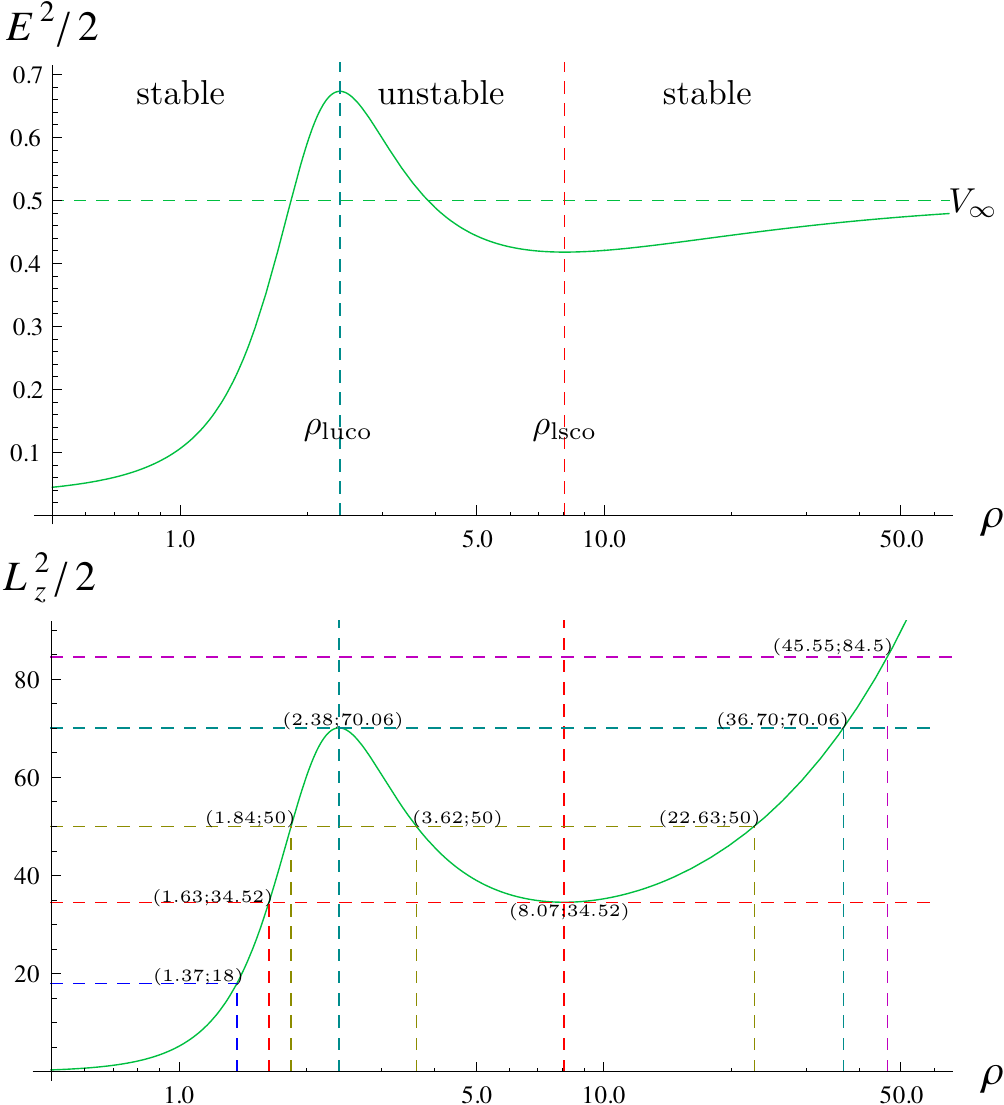}
  \caption{(color online) The functions $E^2(\rho)/2$ and $L_z^2(\rho)/2$ for $M=1.5$ as an example of the interval $M_*<M<\bar{M}$. There can be one, two, or three circular orbits. The area for $L_z^2/2$ with three circular orbits is bounded (see also Fig.~\ref{pic:MP2MetricTimelikeParticleInEquatorialPlaneEffectivePotential2}). These boundaries are given by the radii $\rho_{\text{lsco}}$ and $\rho_{\text{luco}}$, which follow from solving Eq.~\eqref{equ:MP2MetricExtremalPointsOfMathcalLAndE}.}
  \label{pic:MP2MetricTimelikeParticleInEquatorialPlaneEnergyAndAngularMomentum2}
\end{figure}

\begin{figure}
  \centering
  \includegraphics[width=\standardSize\textwidth]{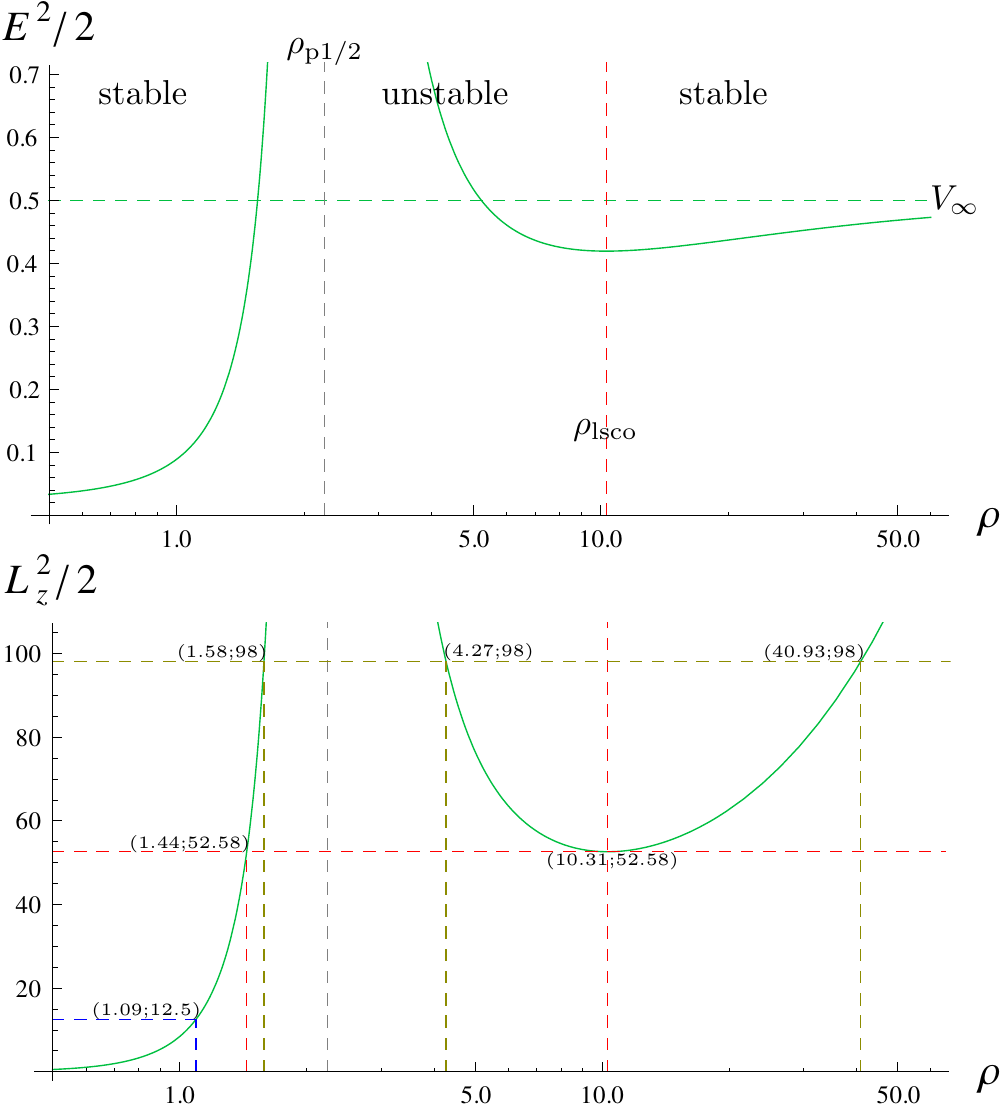}
  \caption{(color online) The functions $E^2(\rho)/2$ and $L_z^2(\rho)/2$ for $M=\bar{M}=\sqrt{27/8}\approx 1.837$. There can be one, two, or three circular orbits. The area for $L_z^2/2$ with three circular orbits no longer has an upper boundary (see also Fig.~\ref{pic:MP2MetricTimelikeParticleInEquatorialPlaneEffectivePotential3}). The singularity is given by the two degenerate photon orbits $\rho_{\text{p}1/2}$ and the local minimum by $\rho_{\text{lsco}}$, which follows from solving Eq.~\eqref{equ:MP2MetricExtremalPointsOfMathcalLAndE}.}
  \label{pic:MP2MetricTimelikeParticleInEquatorialPlaneEnergyAndAngularMomentum3}
\end{figure}

\begin{figure}
  \centering
  \includegraphics[width=\standardSize\textwidth]{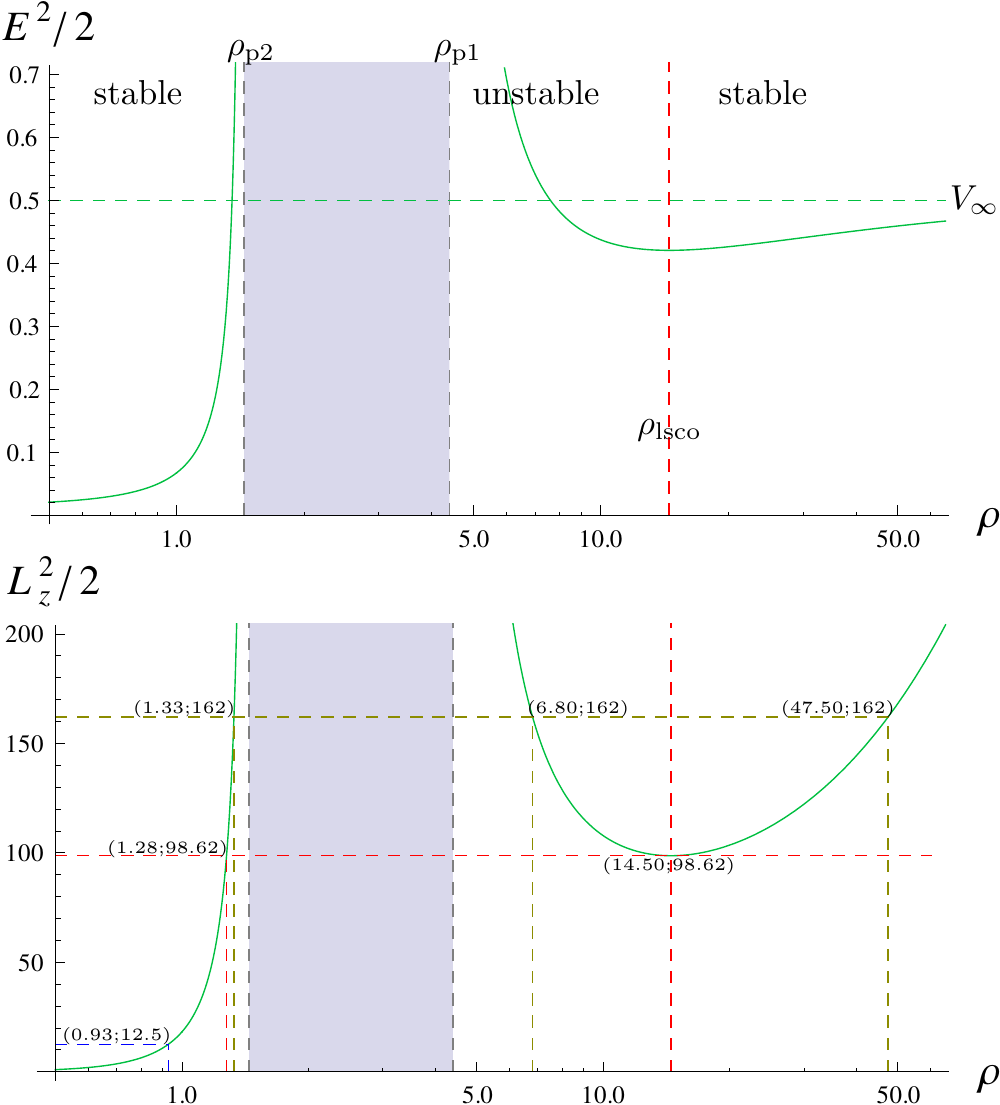}
  \caption{(color online) The functions $E^2(\rho)/2$ and $L_z^2(\rho)/2$ for $M=2.5$ as an example of the interval $M>\bar{M}$. The behavior is qualitatively similar to the case $M=\bar{M}$. The corresponding effective potential is shown in Fig.~\ref{pic:MP2MetricTimelikeParticleInEquatorialPlaneEffectivePotential4}. Between the two photon orbits $\rho_{\text{p}1}$ and $\rho_{\text{p}2}$, there cannot be any circular orbits. The local minimum is given by $\rho_{\text{lsco}}$ and follows from solving Eq.~\eqref{equ:MP2MetricExtremalPointsOfMathcalLAndE}.}
  \label{pic:MP2MetricTimelikeParticleInEquatorialPlaneEnergyAndAngularMomentum4}
\end{figure}

Here, the position of circular orbits depends on $L_z$ in contrast to the lightlike case. In the potential, $L_z$ appears only quadratic; 
%as well as $E$ in Eq.~\eqref{equ:MP2MetricCylindricalCoordinatesEnergyBalanceEquatorialPlane}. Therefore, we define 
%\begin{equation}
%   \mathcal{L} := \frac{L_z^2}{2}, \quad \mathcal{E} := \frac{E^2}{2}.
%\label{equ:MP2MetricDefinitionMathcalL}
%\end{equation}
%
thus, when considering $L_z^2$, we do not have to distinguish between the circumferential direction on the circular geodesic. For the search of extremal points, we have to solve the equation $\partial V_{\text{eff}}/\partial \rho=0$, which can be resolved with respect to $L_z^2/2$. Again, we substitute $\var:=\rho^2+1$ and obtain
\begin{align}
   \frac{L_z^2(\var)}{2} = \ &\frac{M\left[\var^3+4M\var^{5/2}+(4M^2-2)\var^2-8\var^{3/2}\right]}{\var(\var^{3/2}-2M\var+4M)} \notag \\
                      &+\frac{M\left[(1-8M^2)\var+4M\var^{1/2}+4M^2\right]}{\var(\var^{3/2}-2M\var+4M)}.
\label{equ:MP2MetricMathcalLFromX}
\end{align}
Next, we replace $L_z^2/2$ in the effective potential from Eq.~\eqref{equ:MP2MetricCylindricalCoordinatesEnergyBalanceEquatorialPlaneEffectivePotential}. Using $\dot{\rho}=0$, we obtain from Eq.~\eqref{equ:MP2MetricCylindricalCoordinatesEnergyBalanceEquatorialPlane} the expression
\begin{equation}
   \frac{E^2(\var)}{2} = \frac{\var(\var^{3/2}+2M)}{2(\var^{1/2}+2M)^2(\var^{3/2}-2M\var+4M)} .
\label{equ:MP2MetricMathcalEFromX}
\end{equation}
These two functions give the angular momentum and the energy of a particle on a circular orbit with radius $\rho=\sqrt{\var-1}$. They are plotted in Figs.~\ref{pic:MP2MetricTimelikeParticleInEquatorialPlaneEnergyAndAngularMomentum1}--\ref{pic:MP2MetricTimelikeParticleInEquatorialPlaneEnergyAndAngularMomentum4} as functions of $\rho$ for different masses $M$. The functions have singularities at the photon orbits $\rho_{\text{p}1/2}$ and are negative in between. The angular momentum has the root $\var=1$, which corresponds to the radius $\rho=0$. This value represents the circular orbit at the center point of the two masses with vanishing radius and coincides with the point of equilibrium from Eq.~\eqref{equ:pointOfEquilibrium}.
%\begin{figure}
%  \centering
%  %\includegraphics[width=0.48\textwidth]{timelikeEnergyAndAngularMom1}
%  \includegraphics[width=\standardSize\textwidth]{wuensch_figure11}
%  \caption{(color online) The functions $E^2(\rho)/2$ and $L_z^2(\rho)/2$ for $M=1$ as an example of the interval $0<M\leq M_*\approx 1.029$ [see Eq.~\eqref{equ:MP2MetricDefinitionOfMStar}]. For all $L_z^2/2$, there is one circular orbit (see also Fig.~\ref{pic:MP2MetricTimelikeParticleInEquatorialPlaneEffectivePotential1}). For $M=M_*$, there would be a reflection point in the diagrams. In this and the following figures, the stability refers to the radial direction.}
%  \label{pic:MP2MetricTimelikeParticleInEquatorialPlaneEnergyAndAngularMomentum1}
%\end{figure}
%

From Figs.~\ref{pic:MP2MetricTimelikeParticleInEquatorialPlaneEnergyAndAngularMomentum1}--\ref{pic:MP2MetricTimelikeParticleInEquatorialPlaneEnergyAndAngularMomentum4} it is obvious that there is a different number of circular orbits for different values of $M$ and $L_z^2/2$. For an exact analysis, the extremal points have to be calculated. Both $\partial L_z^2/\partial \var=0$ and $\partial E^2/\partial \var=0$ lead to the condition
\begin{equation}
   f(\var) := \var^3-6M\var^{5/2}+3\var^2+22M\var^{3/2}+16 M^2 = 0.
\label{equ:MP2MetricExtremalPointsOfMathcalLAndE}
\end{equation}  
Additionally, as in the lightlike case, we will analyze the stability of the circular orbits in the $\rho$ direction. To do so, we consider $\partial^2 V_{\text{eff}}/\partial \rho^2=0$ and again use the substitution $\var=\rho^2+1$ and Eq.~\eqref{equ:MP2MetricMathcalLFromX}. The result is also the expression from Eq.~\eqref{equ:MP2MetricExtremalPointsOfMathcalLAndE}. Thus, the extremal points of $L_z^2/2$ and $E^2/2$ are linked directly to the reflection points of $V_{\text{eff}}(\rho)$, and we obtain the radial stability of a circular orbit from the sign of their derivative (such as those shown in the plots of $E^2/2$ in Figs.~\ref{pic:MP2MetricTimelikeParticleInEquatorialPlaneEnergyAndAngularMomentum1}--\ref{pic:MP2MetricTimelikeParticleInEquatorialPlaneEnergyAndAngularMomentum4}).
\begin{figure}[b]
  \centering
  \includegraphics[width=\standardSize\textwidth]{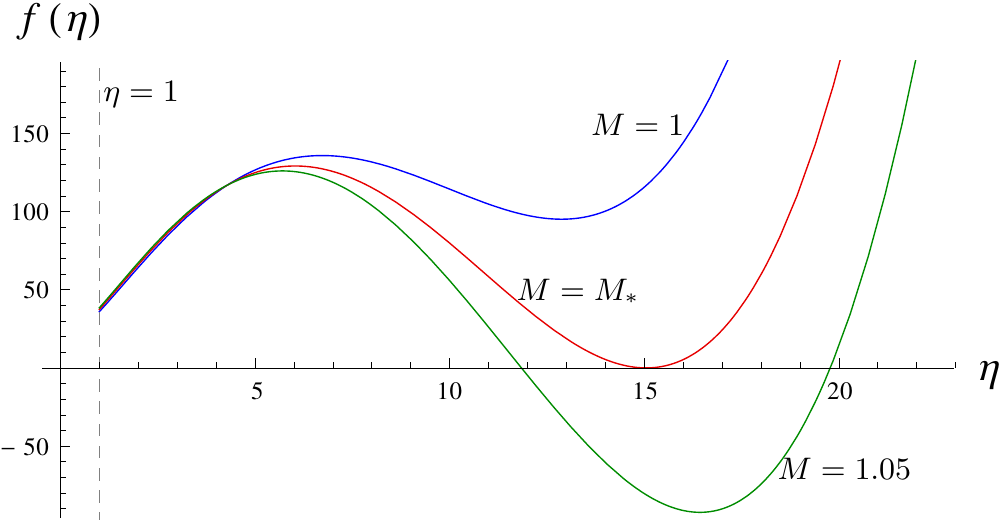}
  \caption{(color online) The function $f(\var)$ from Eq.~\eqref{equ:MP2MetricExtremalPointsOfMathcalLAndE} for different mass values. The number of roots depends on $M$.}
  \label{pic:MP2MetricTimelikeParticleInEquatorialPlanePictureToMStar}
\end{figure}

\begin{figure}
  \centering
  \includegraphics[width=\standardSize\textwidth]{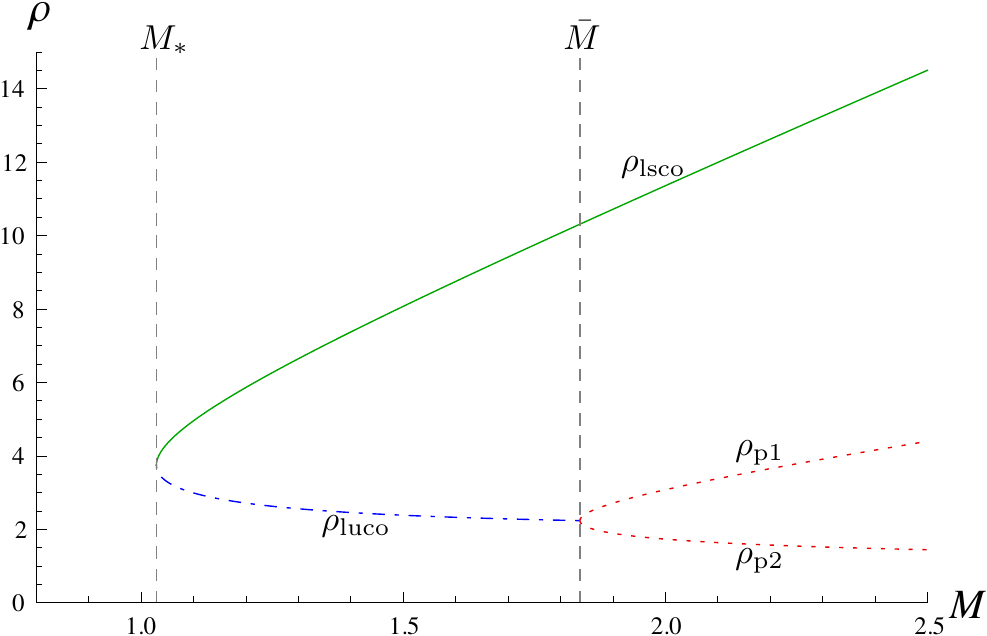}
  \caption{(color online) Numerical solution of Eq.~\eqref{equ:MP2MetricExtremalPointsOfMathcalLAndE}. Solutions exist for $M\geq M_{*}$. For $M\geq\bar{M}$, the dash-dotted branch would lie between the two photon orbits $\rho_{\text{p}1/2}$ (see also Fig.~\ref{pic:MP2MetricPhotonOrbitsInEquatorialPlane}) but would have no physical meaning.}
  \label{pic:MP2MetricTimelikeParticleInEquatorialPlaneLastStableAndUnstableCircularOrbit}
\end{figure}

In Fig.~\ref{pic:MP2MetricTimelikeParticleInEquatorialPlanePictureToMStar}, we present $f(\var)$ of Eq.~\eqref{equ:MP2MetricExtremalPointsOfMathcalLAndE} for several values of $M$. The existence of roots depends on the explicit value of $M$. As shown in Fig.~\ref{pic:MP2MetricTimelikeParticleInEquatorialPlanePictureToMStar}, there is a characteristic value $M_*$ that divides the $M$ range in an interval with no roots of $f(\var)$ and in an interval with two roots. For $M=M_*$, there is exactly one root. This mass $M_*$ follows from $f(\var)=0$ and $f'(\var)=0$ and is given by
\begin{equation}
   M_* = \frac{(13+\sqrt{129})\sqrt{710+70\sqrt{129}}}{50(7+\sqrt{129})} \approx 1.02949.
\label{equ:MP2MetricDefinitionOfMStar}
\end{equation} 
Equation~\eqref{equ:MP2MetricExtremalPointsOfMathcalLAndE} is discussed numerically. If roots exist ($M \geq M_*$), they are called $\rho_{\text{lsco}}$ and $\rho_{\text{luco}}$, and are shown in Fig.~\ref{pic:MP2MetricTimelikeParticleInEquatorialPlaneLastStableAndUnstableCircularOrbit}. The subscript {\textquotedblleft}lsco{\textquotedblright} (last stable circular orbit) is motivated by the analogy to the corresponding radius known from the Schwarzschild~\cite{RindlerRelativity} or the RN spacetime~\cite{Ruffini2011a}. Here, however, we refer only to the radial stability. Furthermore, $\rho_{\text{lsco}}$ is the boundary where the radial stability character changes (compare with Figs.~\ref{pic:MP2MetricTimelikeParticleInEquatorialPlaneEnergyAndAngularMomentum2}--\ref{pic:MP2MetricTimelikeParticleInEquatorialPlaneEnergyAndAngularMomentum4}). For $\rho\leq\rho_{\text{lsco}}$, the circular orbits are unstable, and for $\rho>\rho_{\text{lsco}}$, they are radially stable. The radius $\rho_{\text{luco}}$ exhibits the opposite behavior, and for this reason we use the subscript {\textquotedblleft}luco{\textquotedblright} (last unstable circular orbit).

In fact, for $\rho<\rho_{\text{luco}}$, there again appears a domain with radially stable orbits, and the innermost stable circular orbit is given by $\var=1$, which corresponds to $\rho=0$. Nevertheless, we will call the local minimum of $E^2/2$ and $L_z^2/2$ the last stable circular orbit. For $M\geq \bar{M}$, the numerical solution of Eq.~\eqref{equ:MP2MetricExtremalPointsOfMathcalLAndE} corresponding to $\rho_{\text{luco}}$ is between the two photon orbits $\rho_{\text{p}1}$ and $\rho_{\text{p}2}$; thus $\rho_{\text{luco}}$ does not exist in this $M$ range. The radii $\rho_{\text{lsco}}$ and $\rho_{\text{luco}}$ are also marked in Figs.~\ref{pic:MP2MetricTimelikeParticleInEquatorialPlaneEffectivePotential2}--\ref{pic:MP2MetricTimelikeParticleInEquatorialPlaneEffectivePotential4} and Figs.~\ref{pic:MP2MetricTimelikeParticleInEquatorialPlaneEnergyAndAngularMomentum2}--\ref{pic:MP2MetricTimelikeParticleInEquatorialPlaneEnergyAndAngularMomentum4}.
\begin{figure}[b]
  \centering
  \includegraphics[width=\standardSize\textwidth]{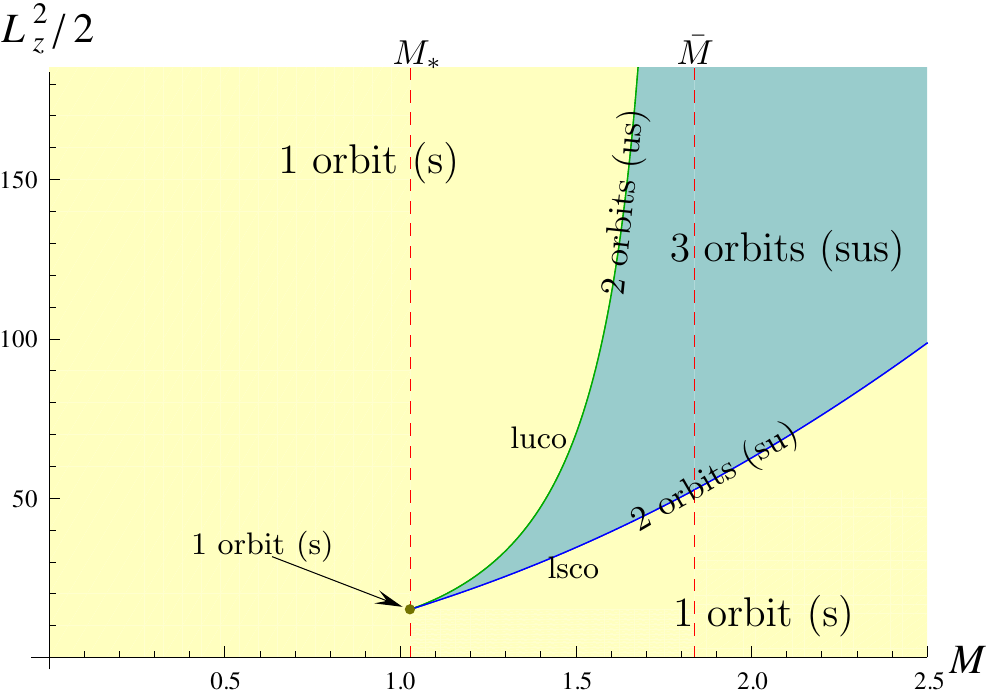}
  \caption{(color online) Number of circular timelike orbits with dependence on the parameters $M$ and $L_z^2/2$. The different regions are colored uniformly. In parentheses, we indicate whether the orbits are radially stable (s) or unstable (u). The order is given by their appearance with increasing $\rho$ in the effective potential from Eq.~\eqref{equ:MP2MetricCylindricalCoordinatesEnergyBalanceEquatorialPlaneEffectivePotential}.}
  \label{pic:MP2MetricTimelikeParticleInEquatorialPlaneNumberOfCircularOrbitsDependingOnMAndMathcalL}
\end{figure}

The regions with different numbers of circular orbits are bounded by $\rho_{\text{lsco}}$ and $\rho_{\text{luco}}$ or $\var_{\text{lsco}}$ and $\var_{\text{luco}}$, respectively. Inserting these numerical values into Eq.~\eqref{equ:MP2MetricMathcalLFromX}, we obtain corresponding boundary lines in the $(M,L_z^2/2)$ plane. These are displayed in Fig.~\ref{pic:MP2MetricTimelikeParticleInEquatorialPlaneNumberOfCircularOrbitsDependingOnMAndMathcalL}. The figure shows the number of circular timelike orbits with dependence on the parameters $M$ and $L_z^2/2$ of the effective potential from Eq.~\eqref{equ:MP2MetricCylindricalCoordinatesEnergyBalanceEquatorialPlaneEffectivePotential}.

\begin{figure}
  \centering
  \includegraphics[width=\standardSize\textwidth]{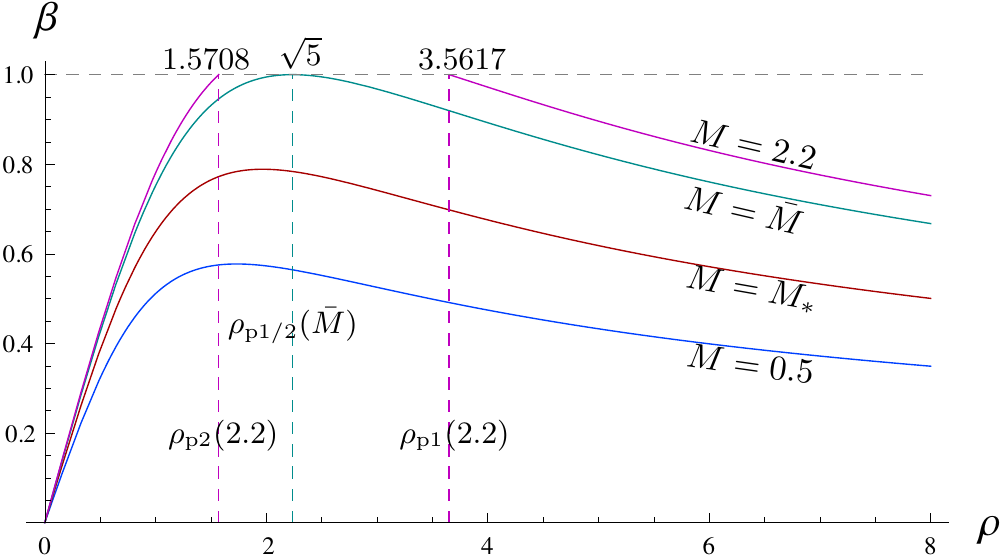}
  \caption{(color online) Local velocity $\beta(\rho)$ on a circular orbit with radius $\rho$. This velocity vanishes for $\rho=0$, and becomes $\beta=1$ at the photon orbits $\rho_{\text{p}1/2}$.}
  \label{pic:MP2MetricTimelikeParticleInEquatorialPlaneLocalVelocityOnCircularOrbit}
\end{figure}

To remain on a circular orbit a particular velocity is necessary. From Eq.~\eqref{equ:MP2MetricCylindricalCoordinatesConstantsOfMotionRespectiveLocalTetrad}, we obtain the angular momentum $L_z$ of a timelike particle. We restrict ourselves to circular orbits ($\xi=\pi/2$) in the equatorial plane ($\chi=\pi/2$) and obtain
\begin{equation}
   \frac{L_z^2}{2} = \frac{1}{2} (\var-1) \left( 1+\frac{2M}{\sqrt{\var}} \right)^2 (\gamma^2-1) ,
\label{equ:MP2MetricTimelikeParticleInEquatorialPlaneAngularMomentumOfTimelikeOrbitRespectivelyTetrad}
\end{equation} 
with $\var:=\rho^2+1$. Comparing this relation with Eq.~\eqref{equ:MP2MetricMathcalLFromX} leads to the necessary local velocity $\beta(\var)$ on the circular orbit with radius $\rho(\var)$ depending on the parameter $M$. After resubstituting, we obtain
\begin{equation}
   \beta(\rho) = \sqrt{\frac{2M\rho^2}{2M+(\rho^2+1)^{3/2}}}.
\label{equ:MP2MetricTimelikeParticleInEquatorialPlaneLocalVelocityOnCircularOrbit}
\end{equation}
This function is shown for several masses $M$ in Fig.~\ref{pic:MP2MetricTimelikeParticleInEquatorialPlaneLocalVelocityOnCircularOrbit}. 
At $\rho=0$, we have $\beta=0$, which again
%Always we get $\beta(\rho=0)=0$. This again 
is the degenerated circular orbit at the point of equilibrium. At the photon orbits $\rho_{\text{p}1/2}$, the velocity is $\beta=1$, and in between it becomes $\beta>1$. 
%Thus in the range $\rho_{\text{p}2}\leq\rho\leq\rho_{\text{p}1}$, no timelike circular orbits exist. 
Thus, there are no timelike circular orbits in the range $\rho_{\text{p}2}\leq\rho\leq\rho_{\text{p}1}$. 
This agrees with $L_z^2/2$ and $E^2/2$ from Eqs.~\eqref{equ:MP2MetricMathcalLFromX} and~\eqref{equ:MP2MetricMathcalEFromX}, which are singular at the photon orbits and undefined in between.
%

%% ------------------------------------------------------------------------
%%         Circular orbits outside the equatorial plane
%% ------------------------------------------------------------------------
\section{Circular orbits outside the equatorial plane}   \label{sec:circularOrbitsOutsideEquatorialPlane}

In Sec.~\ref{sec:circularOrbitsInEquatorialPlane}, we have shown, by means of Eq.~\eqref{equ:MP2MetricGeneralConditionForCircularOrbits}, that lightlike and timelike circular orbits only exist for ${z\in(-1;+1)}$. For $z=0$, this equation was independent of $\rho$ and implied $M_1=M_2$. For $0<|z|<1$, Eq.~\eqref{equ:MP2MetricGeneralConditionForCircularOrbits} can be rearranged to
\begin{equation}
   \rho^2 = \frac{4z}{1-\left(\frac{M_1}{M_2}\right)^{2/3} \left(\frac{1+z}{1-z}\right)^{2/3}} - (z+1)^2,
\label{equ:MP2MetricCircularOrbitsOutsideEquatorialPlaneRhoSquaredFromZ}
\end{equation}
which now shows a direct dependence on the radius $\rho$ and the height $z$ for the motion with $\dot{z}=0$. The existence of circular orbits outside the equatorial plane can be understood classically, at least for massive particles, by analyzing the gravitational forces; see Appendix~\ref{app:classicalAnalogon}. Photon orbits are a purely relativistic effect.
\begin{figure*}
  \centering
  \includegraphics[width=0.65\textwidth]{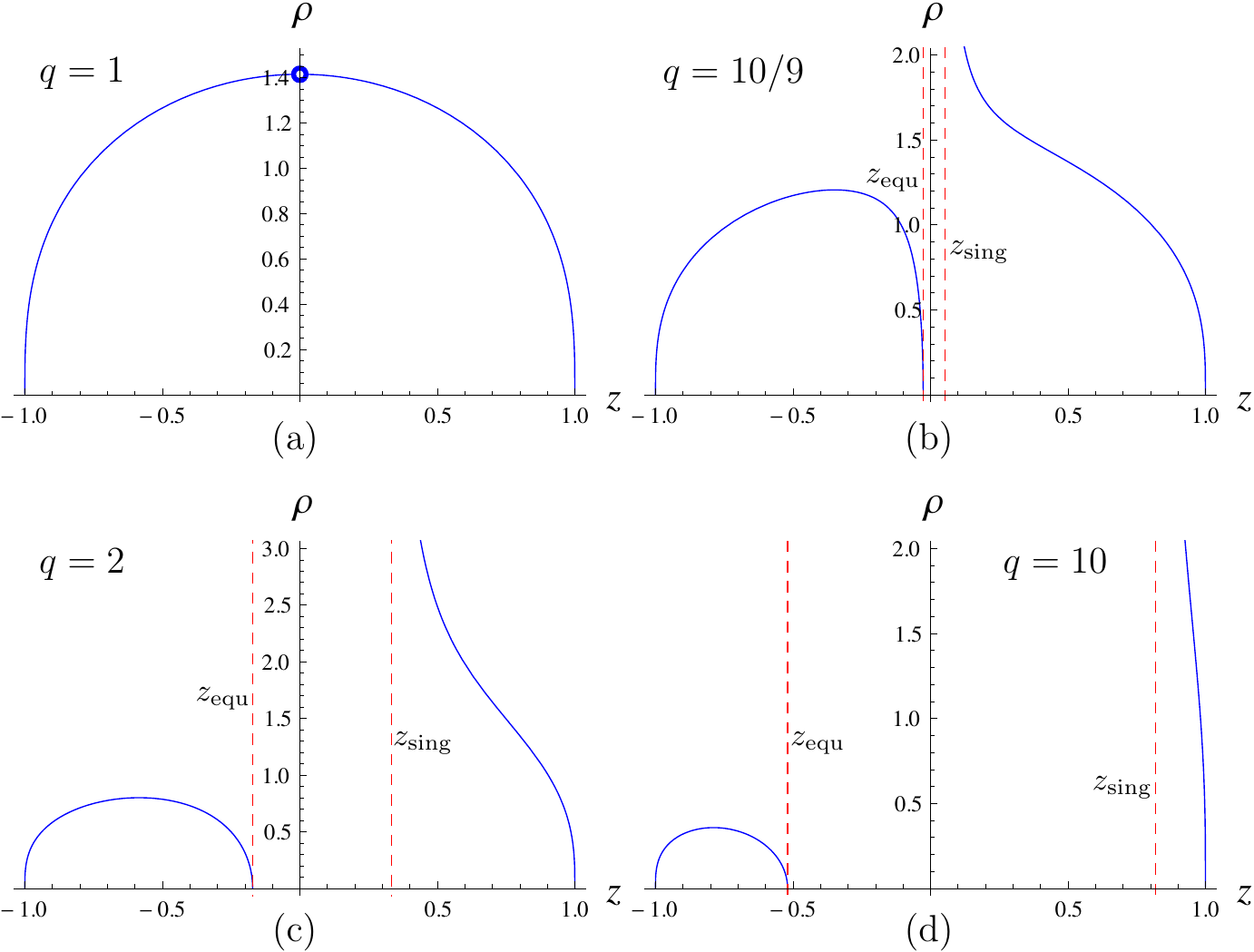}
  \caption{(color online) Radius $\rho(z)$ of a circular orbit at height $z$ above the equatorial plane for several mass ratios $q$; see Eq.~\eqref{equ:MP2MetricCircularOrbitsOutsideEquatorialPlaneRhoSquaredFromZ}. In the definition range $z\in (-1;+1)$, there is an additionally forbidden domain for $q\neq 1$ between $z_{\text{equ}}$ and $z_{\text{sing}}$.}
  \label{pic:rhoFromZForSeveralQ}
\end{figure*}

The radius $\rho(z)$ depends only on the mass ratio ${q:=M_1/M_2}$, and is shown as a function of $z$ in Fig.~\ref{pic:rhoFromZForSeveralQ} for several values of $q$. For equal masses ($q=1$), all heights $z$ with $0<|z|<1$ are possible candidates for circular orbits. In the case of unequal masses ($q\neq 1$), there is a domain without allowed radii ($\rho\geq 0$), which is bounded by the two characteristic heights $z_{\text{equ}}(q)$ and $z_{\text{sing}}(q)$. The boundary $z_{\text{equ}}(q)$ agrees with the point of equilibrium from Eq.~\eqref{equ:pointOfEquilibrium} (see also Fig.~\ref{pic:pointOfEquilibrium}) and results from root-finding of $\rho(z)$.
%\begin{figure*}
%  \centering
%  %\includegraphics[width=0.48\textwidth]{rhoFromZForSeveralQ}
%  \includegraphics[width=0.7\textwidth]{wuensch_figure19}
%  \caption{(color online) Radius $\rho(z)$ of a circular orbit at height $z$ above the equatorial plane for several mass ratios $q$; see Eq.~\eqref{equ:MP2MetricCircularOrbitsOutsideEquatorialPlaneRhoSquaredFromZ}. In the definition range $z\in (-1;+1)$, there is an additionally forbidden domain for $q\neq 1$ between $z_{\text{equ}}$ and $z_{\text{sing}}$.}
%  \label{pic:rhoFromZForSeveralQ}
%\end{figure*}
%
The singularity $z_{\text{sing}}(q)$ follows from the root of the denominator in Eq.~\eqref{equ:MP2MetricCircularOrbitsOutsideEquatorialPlaneRhoSquaredFromZ} and is given by
\begin{equation}
   z_{\text{sing}}(q) = \frac{q-1}{q+1},  
\label{equ:singularityPoint}
\end{equation}
a quantity already shown in Fig.~\ref{pic:pointOfEquilibrium}. The more the ratio $q$ of the two masses differs from unity, the more extended the interval between $z_{\text{equ}}$ and $z_{\text{sing}}$ becomes.

Equation~\eqref{equ:MP2MetricCircularOrbitsOutsideEquatorialPlaneRhoSquaredFromZ} follows from $\partial V_{\text{eff}}/\partial z=0$ with the effective potential~\eqref{equ:MP2MetricCylindricalCoordinatesEnergyBalanceEffectivePotential}. For a circular orbit, we also have to fulfill $\partial V_{\text{eff}}/\partial \rho=0$.

\subsection{Null geodesics}   \label{sec:lightlikeCircularOrbitsOutsideEquatorialPlane}

\begin{figure}[b]
  \centering
  \includegraphics[width=\standardSize\textwidth]{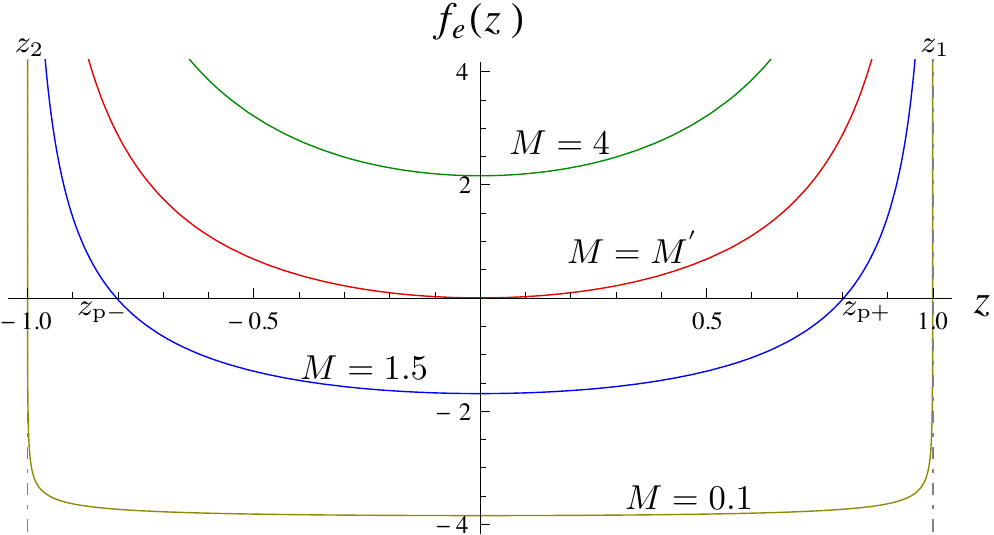}
  \caption{(color online) The function $f_{\text{e}}(z)$ from Eq.~\eqref{equ:MP2MetricCircularOrbitsOutsideEquatorialPlaneEqualMassesUglyCondition} for several masses $M$. The heights of the photon orbits are given by the roots. Their number depends on the concrete value of $M$.}
  \label{pic:MP2MetricFeFromZForEqualMasses}
\end{figure}

First, we consider the case in which both masses are equal, $M:=M_1=M_2$. The prerequisite $\partial V_{\text{eff}}/\partial \rho=0$ leads to
\begin{equation}
   U + 2\rho \frac{\partial U}{\partial \rho} = 0.
\label{equ:MP2MetricDerivativeAfterRhoEqualizingZero}
\end{equation}
Using $U(\rho,z)$ from Eq.~\eqref{equ:MP2MetricUFunktionCylindricalCoordinates} and $\rho^2(z)$ from Eq.~\eqref{equ:MP2MetricCircularOrbitsOutsideEquatorialPlaneRhoSquaredFromZ}, the expression~\eqref{equ:MP2MetricDerivativeAfterRhoEqualizingZero} yields the root-finding problem
\begin{align}
   f_{\text{e}}(z):=&\left[4A_{\text{e}}(z)-(1+z)^2\right] \left( \frac{M}{\left[A_{\text{e}}(z)\right]^{\frac{3}{2}}}+\frac{M}{\left[A_{\text{e}}(z)-z\right]^{\frac{3}{2}}} \right) \notag \\
   &-2 \left( 2+ \frac{M}{\left[A_{\text{e}}(z)\right]^{\frac{1}{2}}}+\frac{M}{\left[A_{\text{e}}(z)-z\right]^{\frac{1}{2}}}\right) = 0  
\label{equ:MP2MetricCircularOrbitsOutsideEquatorialPlaneEqualMassesUglyCondition}
\end{align}
with
\begin{equation}
   A_{\text{e}}(z) = \frac{z}{1-\left(\frac{1-z}{1+z}\right)^{2/3}}.
\label{equ:MP2MetricAbreviationAeFromZ}
\end{equation}
The index {\textquotedblleft}e{\textquotedblright} designates the case of equal masses. Figure~\ref{pic:MP2MetricFeFromZForEqualMasses} shows $f_{\text{e}}(z)$ for several $M$.
In the range ${0<M<M^{'}}$, with a characteristic mass $M^{'}$, there are always two circular orbits at heights $z_{\text{p}\pm}$. The numerical value of $M^{'}$ is given by
\begin{equation}
   M^{'} \approx 2.598076.
\label{equ:MP2MetricDefinitionMLine}
\end{equation}
\begin{figure*}
  \centering
  \includegraphics[width=0.9\textwidth]{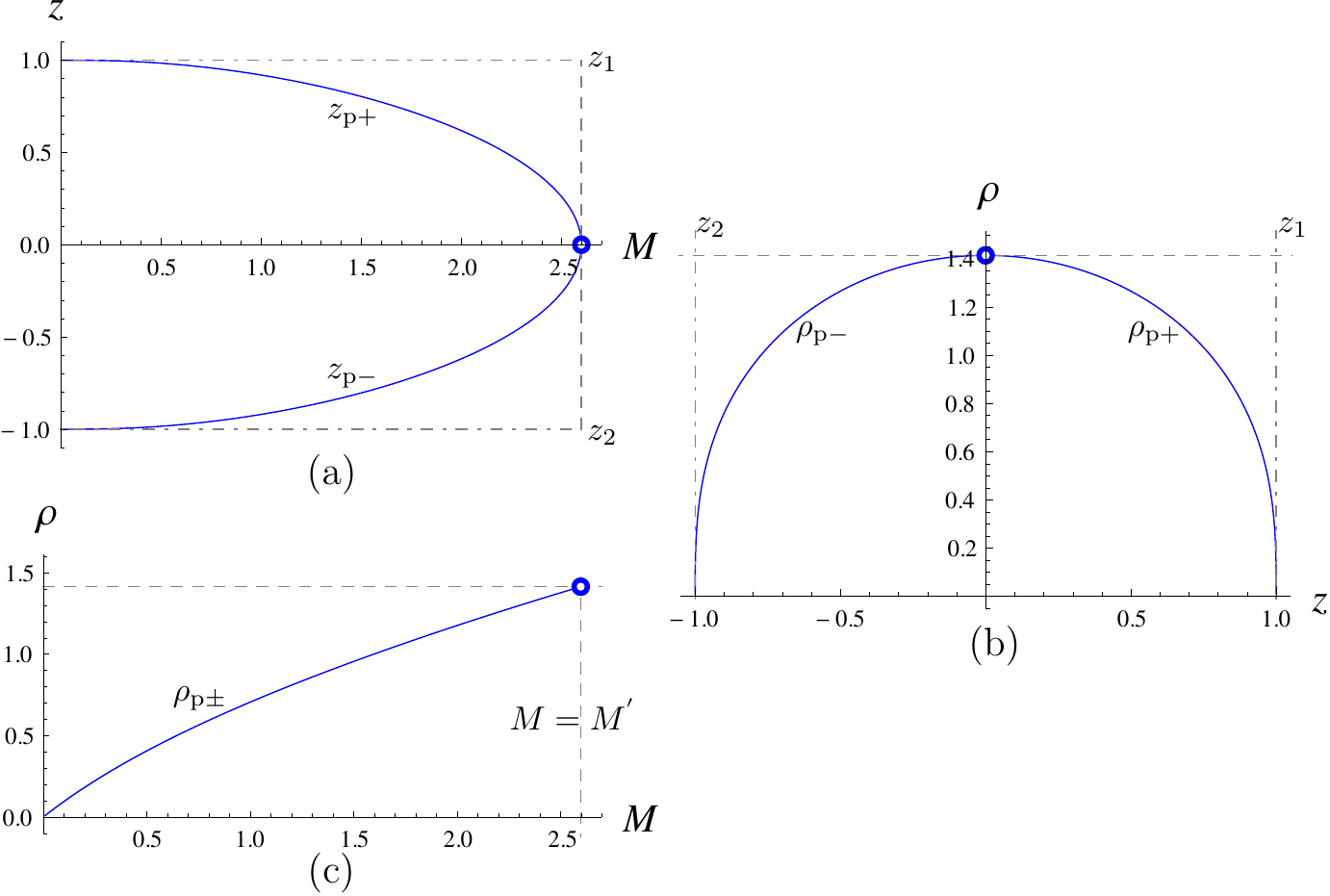}
  \caption{(color online) Numerical analysis of Eq.~\eqref{equ:MP2MetricCircularOrbitsOutsideEquatorialPlaneEqualMassesUglyCondition}. Shown are (a) the heights $z_{\text{p}\pm}$ and the corresponding radii $\rho_{\text{p}\pm}$ obtained from Eq.~\eqref{equ:MP2MetricCircularOrbitsOutsideEquatorialPlaneRhoSquaredFromZ} depending on (b) the heights $z$ and (c) the mass $M$. Because of the equatorial symmetry, we have $z_{\text{p}+}=-z_{\text{p}-}$ and $\rho_{\text{p}+}=\rho_{\text{p}-}$.}
  \label{pic:MP2MetricPhotonOrbitsOutsideEquatorialPlaneForEqualMasses}
\end{figure*}

\noindent
For $M>M^{'}$, there are no intersections with the abscissa and no circular orbits. In the limiting case $M=M^{'}$, the only root is $z_{\text{p}\pm}=0$, which here does not lie in the range of definition. The results of the numerical analysis of Eq.~\eqref{equ:MP2MetricCircularOrbitsOutsideEquatorialPlaneEqualMassesUglyCondition} are shown in Fig.~\ref{pic:MP2MetricPhotonOrbitsOutsideEquatorialPlaneForEqualMasses}. The corresponding radii $\rho_{\text{p}\pm}$ follow from Eq.~\eqref{equ:MP2MetricCircularOrbitsOutsideEquatorialPlaneRhoSquaredFromZ}. For each $z$ with ${0<|z|<1}$, there is exactly one $M\in (0,M^{'})$ such that $z$ is the height of a photon orbit. This is also confirmed by the comparison of Fig.~\ref{pic:MP2MetricPhotonOrbitsOutsideEquatorialPlaneForEqualMasses}(b) with Fig.~\ref{pic:rhoFromZForSeveralQ}(a).
%\begin{figure*}
%  \centering
%  %\includegraphics[width=0.5\textwidth]{PhotonenOrbitsForEqualMasses}
%  \includegraphics[width=0.74\textwidth]{wuensch_figure21}
%  \caption{(color online) Numerical analysis of Eq.~\eqref{equ:MP2MetricCircularOrbitsOutsideEquatorialPlaneEqualMassesUglyCondition}. Shown are (a) the heights $z_{\text{p}\pm}$ and the corresponding radii $\rho_{\text{p}\pm}$ obtained from Eq.~\eqref{equ:MP2MetricCircularOrbitsOutsideEquatorialPlaneRhoSquaredFromZ} depending on (b) the heights $z$ and (c) the mass $M$. Because of the equatorial symmetry, we have $z_{\text{p}+}=-z_{\text{p}-}$ and $\rho_{\text{p}+}=\rho_{\text{p}-}$.}
%  \label{pic:MP2MetricPhotonOrbitsOutsideEquatorialPlaneForEqualMasses}
%\end{figure*}
%

\begin{figure*}
  \centering
  \includegraphics[width=0.9\textwidth]{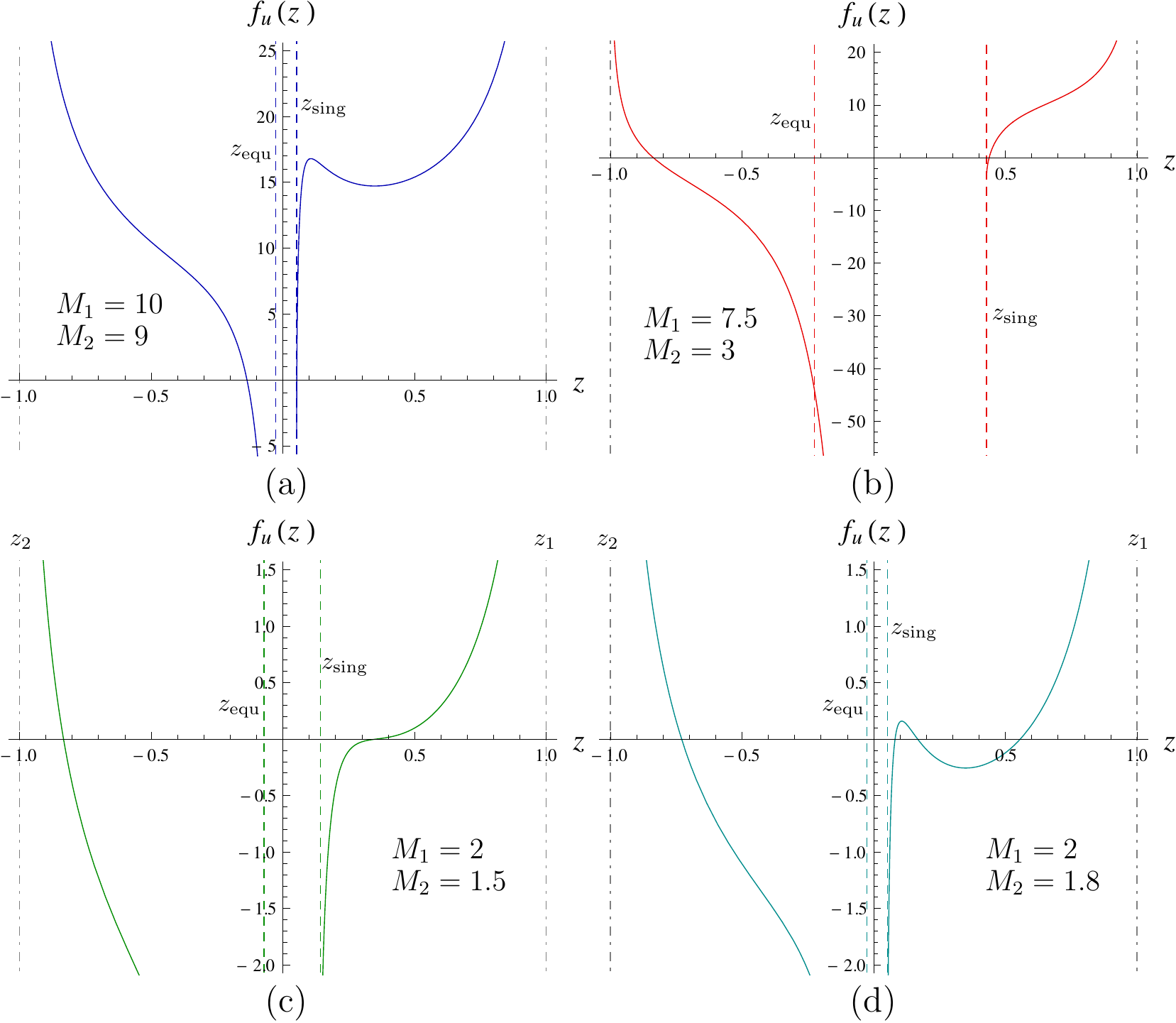}
  \caption{(color online) The function $f_{\text{u}}(z)$ from Eq.~\eqref{equ:MP2MetricCircularOrbitsOutsideEquatorialPlaneUnequalMassesUglyCondition} for several combinations of the mass values $M_1$ and $M_2$. The heights of the photon orbits are given by the roots. Their number varies between two and four.}
  \label{pic:MP2MetricFuFromZForUnequalMasses}
\end{figure*}

For unequal masses $M_1$ and $M_2$, we find the same equation~\eqref{equ:MP2MetricCircularOrbitsOutsideEquatorialPlaneEqualMassesUglyCondition}, but with
\begin{equation}
   A_{\text{u}}\left(z,\frac{M_1}{M_2}\right) = \frac{z}{1-\left(\frac{M_1}{M_2}\right)^{2/3}\left(\frac{1-z}{1+z}\right)^{2/3}},
\label{equ:MP2MetricAbreviationAuFromZ}
\end{equation}
instead of $A_{\text{e}}(z)$ from Eq.~\eqref{equ:MP2MetricAbreviationAeFromZ}. In this case, we will abbreviate the root-finding problem with 
\begin{equation}
   f_{\text{u}}(z)=0.
\label{equ:MP2MetricCircularOrbitsOutsideEquatorialPlaneUnequalMassesUglyCondition}
\end{equation}
The solutions will be called $z_{\text{p}i}$ with an enumerating index $i$. The corresponding radii are $\rho_{\text{p}i}$. Figure~\ref{pic:MP2MetricFuFromZForUnequalMasses} shows $f_{\text{u}}(z)$ for several significant combinations of $M_1$ and $M_2$. The function depends on both masses, not only on their ratio as $\rho(z)$ from Eq.~\eqref{equ:MP2MetricCircularOrbitsOutsideEquatorialPlaneRhoSquaredFromZ}. Because of the symmetry we can restrict ourselves to $M_1>M_2$ at first; thus, we have $z_{\text{equ}}<0$ and $z_{\text{sing}}>0$, without loss of generality.

For $f_{\text{u}}(z)$, we find the following limiting values:
\begin{align}
\begin{aligned}
   &\lim_{z\rightarrow z_{\text{sing}+}}f_{\text{u}}(z) = -4, \ \ \lim_{z\rightarrow 0-}\ \ f_{\text{u}}(z) = -\infty , \\
   &\ \ \lim_{z\rightarrow \pm 1}f_{\text{u}}(z) = \infty.
\end{aligned}
\label{equ:MP2MetricFuFromZForUnequalMassesLimitingValues}
\end{align}
Consequently, there are at least two photon orbits. In the interval $(-1;0)$, the function is monotonic but not necessarily for $z\in (z_{\text{sing}};+1)$. Thus, up to four roots can exist. Here, the two-dimensional parameter space and the complexity of Eq.~\eqref{equ:MP2MetricCircularOrbitsOutsideEquatorialPlaneUnequalMassesUglyCondition} make exact analysis difficult. Significant aspects, however, become clear on the basis of the following two examples.

\begin{figure*}
  \centering
  \includegraphics[width=0.8\textwidth]{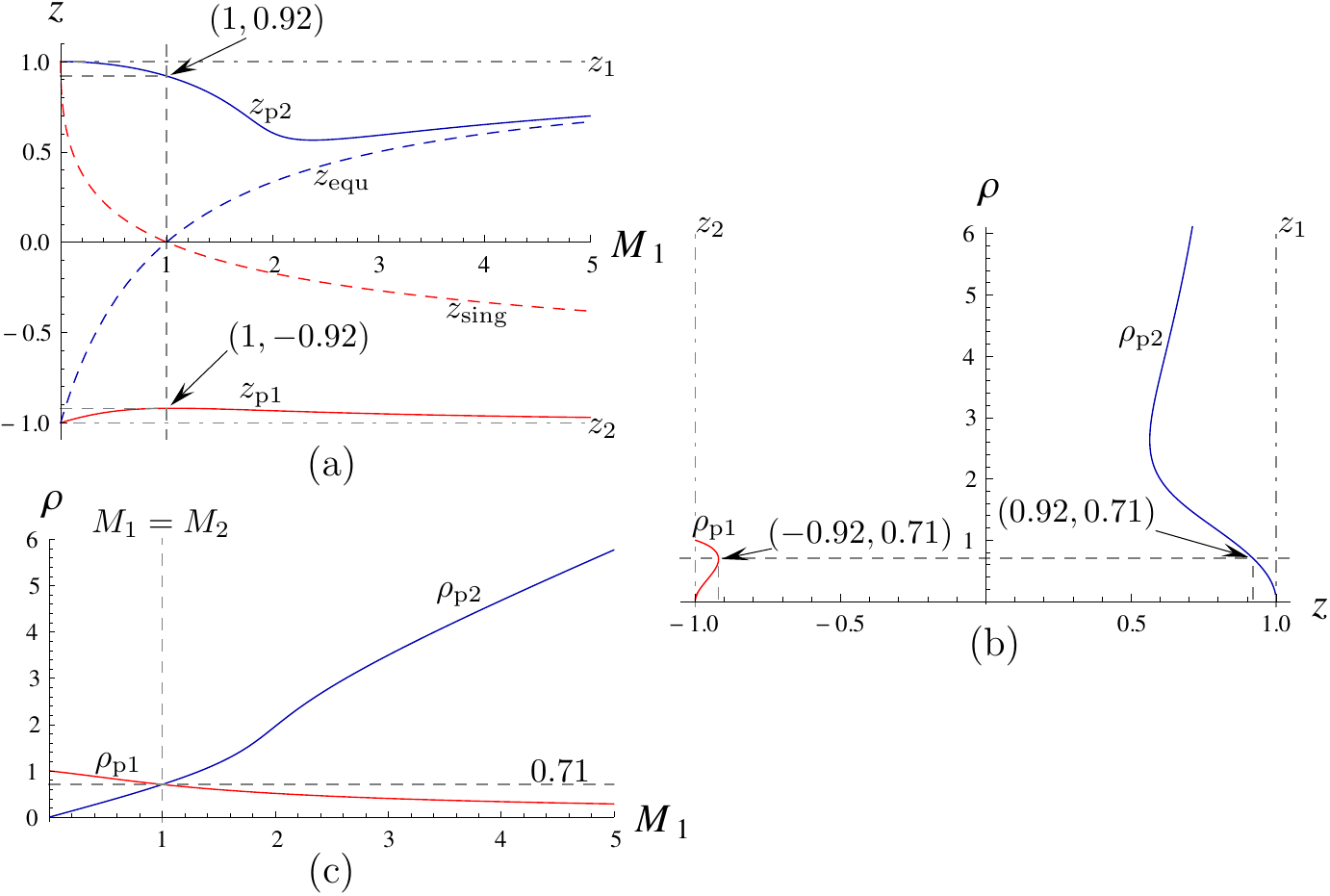}
  \caption{(color online) Numerical analysis of Eq.~\eqref{equ:MP2MetricCircularOrbitsOutsideEquatorialPlaneUnequalMassesUglyCondition} for fixed $M_2=1$ and variable $M_1$. Shown are (a) the heights $z_{\text{p}1/2}$ and the corresponding radii $\rho_{\text{p}1/2}$ obtained from Eq.~\eqref{equ:MP2MetricCircularOrbitsOutsideEquatorialPlaneRhoSquaredFromZ} depending on (b) the heights $z$ and (c) the mass $M_1$. For all $M_1$, there are two photon orbits.}
  \label{pic:MP2MetricPhotonOrbitsOutsideEquatorialPlaneForUnequalMasses1}
\end{figure*}

First, we solve Eq.~\eqref{equ:MP2MetricCircularOrbitsOutsideEquatorialPlaneUnequalMassesUglyCondition} numerically for variable $M_1$, where $M_2=1$ is fixed (see Fig.~\ref{pic:MP2MetricPhotonOrbitsOutsideEquatorialPlaneForUnequalMasses1}). 
There are two photon orbits at $z_{\text{p}1}$ and  $z_{\text{p}2}$ with corresponding radii $\rho_{\text{p}1}$ and  $\rho_{\text{p}2}$. For $M_1=1$, the masses are equal and $z_{\text{p}1}=-z_{\text{p}2}$ in agreement with the already discussed case with equal masses (see also Fig.~\ref{pic:MP2MetricPhotonOrbitsOutsideEquatorialPlaneForEqualMasses}).

As a further example, we fix $M_2=1.8$, and $M_1$ remains variable. The missing monotonicity of $f_{\text{u}}(z)$ for $z\in (z_{\text{sing}};+1)$ for some combinations of $M_1$ and $M_2$ leads to an area of $M_1$ with four photon orbits. Figure~\ref{pic:MP2MetricPhotonOrbitsOutsideEquatorialPlaneForUnequalMasses2} shows the numerical analysis of Eq.~\eqref{equ:MP2MetricCircularOrbitsOutsideEquatorialPlaneUnequalMassesUglyCondition}.
\begin{figure*}
  \centering
  \includegraphics[width=0.8\textwidth]{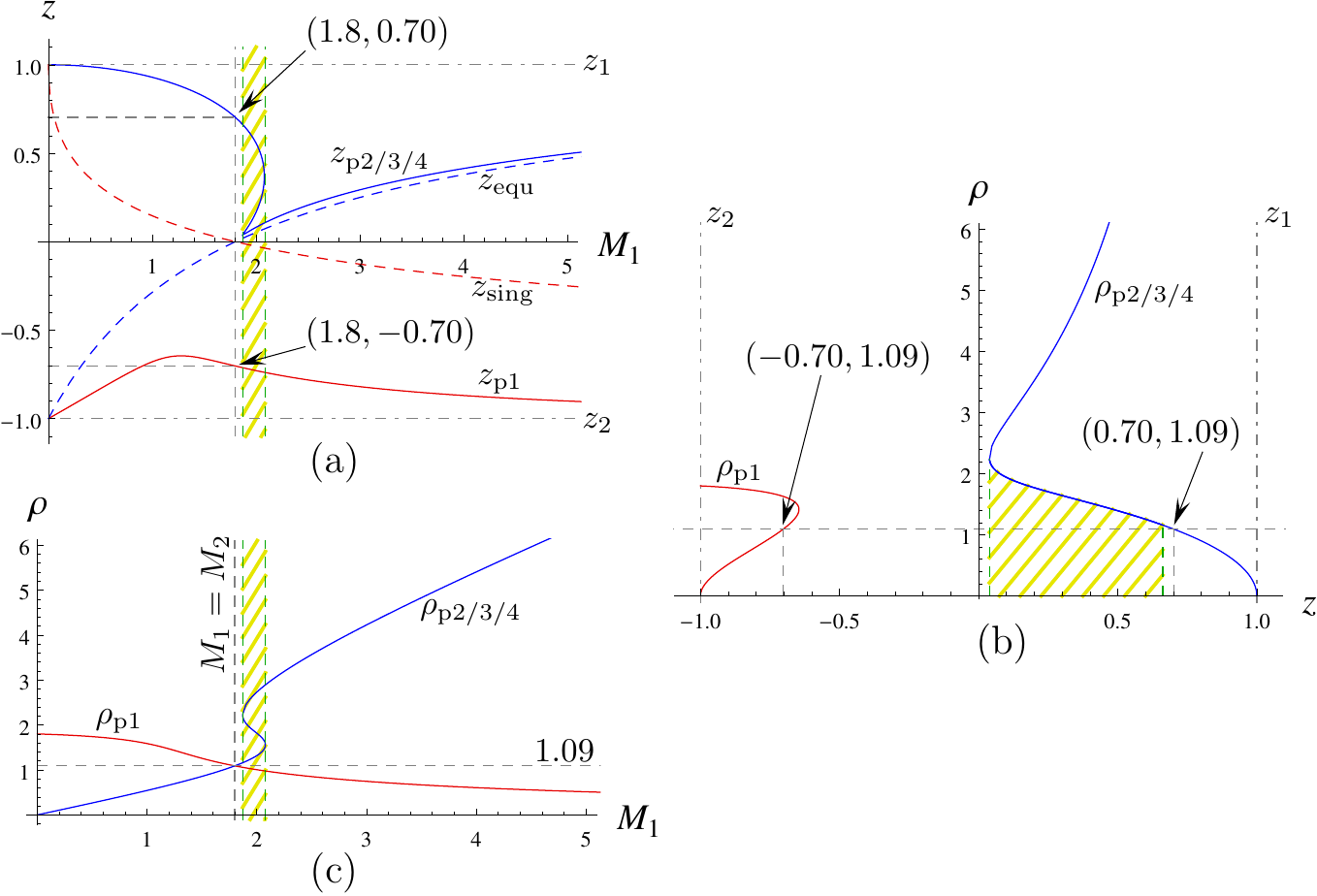}
  \caption{(color online) Numerical analysis of Eq.~\eqref{equ:MP2MetricCircularOrbitsOutsideEquatorialPlaneUnequalMassesUglyCondition} for fixed $M_2=1.8$ and variable $M_1$. Shown are (a) the heights $z_{\text{p}1/2/3/4}$ and the corresponding radii $\rho_{\text{p}1/2/3/4}$ obtained from Eq.~\eqref{equ:MP2MetricCircularOrbitsOutsideEquatorialPlaneRhoSquaredFromZ} depending on (b) the heights $z$ and (c) the mass $M_1$. There is a range of $M_1$ with four photon orbits (shaded domain).}
  \label{pic:MP2MetricPhotonOrbitsOutsideEquatorialPlaneForUnequalMasses2}
\end{figure*}

\subsection{Timelike geodesics}   \label{sec:timelikeCircularOrbitsOutsideEquatorialPlane}

As in the lightlike case, Eq.~\eqref{equ:MP2MetricCircularOrbitsOutsideEquatorialPlaneRhoSquaredFromZ} has to be fulfilled. To stay on a circular orbit the test particle must have the local velocity
\begin{equation}
   \beta = \sqrt{\frac{U}{U+\rho \frac{\partial U}{\partial\rho}}-1};
\label{equ:MP2MetricGeneralConditionForLocelVelocityOnCircularGeodesic}
\end{equation}
see Appendix~\ref{app:generalCircularWorldlineInTheMP2Metric}. Here, we have to use $\rho(z)$ from Eq.~\eqref{equ:MP2MetricCircularOrbitsOutsideEquatorialPlaneRhoSquaredFromZ} and $U(\rho(z),z)$ from Eq.~\eqref{equ:MP2MetricUFunktionCylindricalCoordinates}. For $\beta=1$, Eq.~\eqref{equ:MP2MetricGeneralConditionForLocelVelocityOnCircularGeodesic} consistently simplifies to Eq.~\eqref{equ:MP2MetricDerivativeAfterRhoEqualizingZero}.
% \begin{figure}
%   %\centering
%   % \includegraphics[width=0.42\textwidth]{timelikeOrbitsVelocityEqualMasses}
%   \includegraphics[width=0.45\textwidth]{wuensch_figure25}
%   \caption{(color online) Local velocity $\beta(z)$ from Eq.~\eqref{equ:MP2MetricGeneralConditionForLocelVelocityOnCircularGeodesic} for a circular orbit at height $z$. Shown are plots for equal masses $M:=M_1=M_2$. The domain with existing photon orbits is bounded by the photon orbits discussed in Sec.~\ref{sec:lightlikeCircularOrbitsOutsideEquatorialPlane}.}
%   \label{pic:MP2MetricTimelikeOrbitsOutsideEquatorialPlaneForEqualMasses}
% \end{figure}
% %
As in the lightlike case, we distinguish between the cases with equal and unequal masses. The function $\beta(z)$ is shown in Fig.~\ref{pic:MP2MetricTimelikeOrbitsOutsideEquatorialPlaneForEqualMasses} for some values of $M:=M_1=M_2$. Since $\beta<1$, we can see that there are regions of $z$ where no timelike circular orbits are possible. The boundaries ($\beta=1$) are given by the photon orbits, with $z_{\text{p}\pm}$ calculated in Sec.~\ref{sec:lightlikeCircularOrbitsOutsideEquatorialPlane}. Because of the absence of photon orbits outside the equatorial plane for $M\geq M^{'}$ [Eq.~\eqref{equ:MP2MetricDefinitionMLine}], there are also no timelike circular orbits outside.
% \begin{figure*}
%   \centering
%   %\includegraphics[width=0.5\textwidth]{timelikeOrbitsVelocityUnequalMasses}
%   \includegraphics[width=0.4\textwidth]{wuensch_figure26}
%   \caption{(color online) Local velocity $\beta(z)$ from Eq.~\eqref{equ:MP2MetricGeneralConditionForLocelVelocityOnCircularGeodesic} for a circular orbit at height $z$. Shown are plots for several combinations of the masses $M_1$ and $M_2$. The domain with existing photon orbits is bounded by the photon orbits discussed in Sec.~\ref{sec:lightlikeCircularOrbitsOutsideEquatorialPlane} and the characteristic heights $z_{\text{equ}}$ and $z_{\text{sing}}$.}
%   \label{pic:MP2MetricTimelikeOrbitsOutsideEquatorialPlaneForUnequalMasses}
% \end{figure*}
% %

For the case of unequal masses $M_1$ and $M_2$, the velocity $\beta(z)$ is plotted for several combinations of the mass values in Fig.~\ref{pic:MP2MetricTimelikeOrbitsOutsideEquatorialPlaneForUnequalMasses}. Again, we have an invalid domain between the two characteristic values $z_{\text{equ}}$ and $z_{\text{sing}}$. At the point of equilibrium, we have $\beta(z_{\text{equ}})=0$ and $\rho(z_{\text{equ}})=0$ with $\rho(z)$ according to Eq.~\eqref{equ:MP2MetricCircularOrbitsOutsideEquatorialPlaneRhoSquaredFromZ}, i.e., a circular orbit that is degenerated to a point at $z_{\text{equ}}$, consistent with Sec.~\ref{sec:metric}. The appearance of $z_{\text{sing}}$ with $\beta\rightarrow 0$ for $z\rightarrow z_{\text{sing}+}$ has a special reason. For $z\rightarrow z_{\text{sing}+}$, we have $\rho\rightarrow\infty$; thus, in this limit, there is a circular orbit with infinite radius at this height. The velocity decreases with growing radial distance, which yields a vanishing velocity in the limiting case. These two characteristic heights and the photon orbits delimit the domain with possible timelike circular orbits. The number of photon orbits can vary between two and four (see also Sec.~\ref{sec:lightlikeCircularOrbitsOutsideEquatorialPlane}). 
% \begin{figure}[H]
%   \centering
%   % \includegraphics[width=0.42\textwidth]{timelikeOrbitsVelocityEqualMasses}
%   \includegraphics[width=0.48\textwidth]{wuensch_figure25}
%   \caption{(color online) Local velocity $\beta(z)$ from Eq.~\eqref{equ:MP2MetricGeneralConditionForLocelVelocityOnCircularGeodesic} for a circular orbit at height $z$. Shown are plots for equal masses $M:=M_1=M_2$. The domain with existing photon orbits is bounded by the photon orbits discussed in Sec.~\ref{sec:lightlikeCircularOrbitsOutsideEquatorialPlane}.}
%   \label{pic:MP2MetricTimelikeOrbitsOutsideEquatorialPlaneForEqualMasses}
% \end{figure}
% %

\begin{figure}[H] \setcounter{figure}{26}
  \centering
  \includegraphics[width=0.47\textwidth]{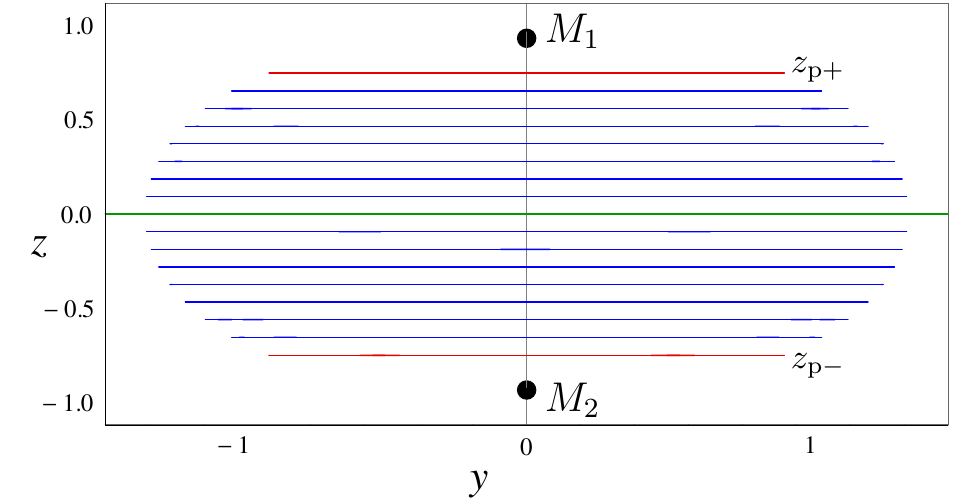}
  \caption{(color online) Edge-on view of circular orbits for $M_1=M_2=1.5$. The photon orbits are located at ${z_{\text{p}\pm}\approx\pm 0.80324}$ with $\rho_{\text{p}\pm}\approx 0.95502$. In the plane ${z=z_{\text{sing}}=z_{\text{equ}}=0}$, there are no photon orbits but an arbitrary number of timelike orbits (compare with Secs.~\ref{sec:lightlikeCircularOrbitsInEquatorialPlane} and~\ref{sec:timelikeCircularOrbitsInEquatorialPlane}).}
  \label{pic:MP2MetricCircularOrbitsOutsideEquatorialPlaneExample1}
\end{figure}
\begin{figure} \setcounter{figure}{24}
  \centering
  \includegraphics[width=0.48\textwidth]{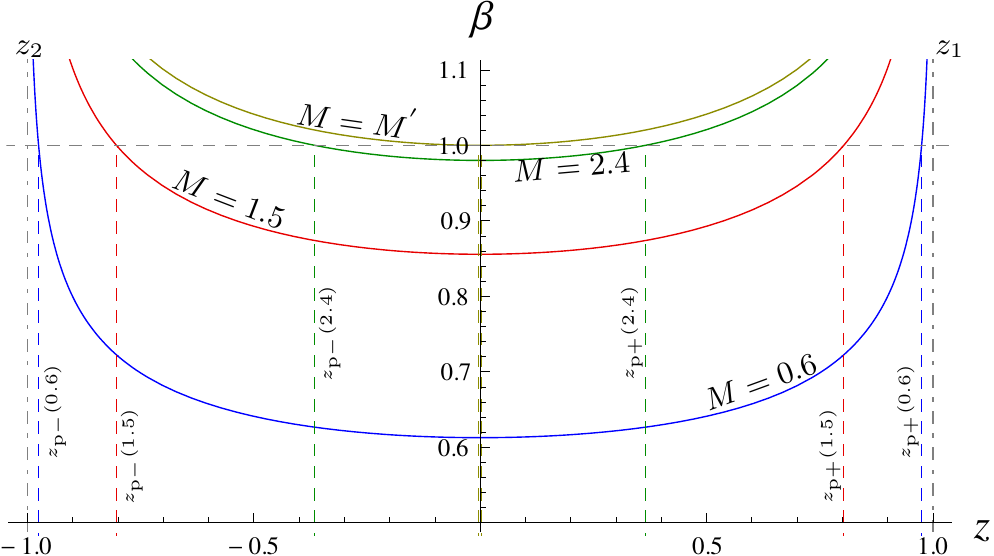}
  \caption{(color online) Local velocity $\beta(z)$ from Eq.~\eqref{equ:MP2MetricGeneralConditionForLocelVelocityOnCircularGeodesic} for a circular orbit at height $z$. Shown are plots for equal masses $M:=M_1=M_2$. The domain with existing photon orbits is bounded by the photon orbits discussed in Sec.~\ref{sec:lightlikeCircularOrbitsOutsideEquatorialPlane}.}
  \label{pic:MP2MetricTimelikeOrbitsOutsideEquatorialPlaneForEqualMasses}
\end{figure}
\onecolumngrid
\ \ \
\begin{figure}[b] \setcounter{figure}{25}
  \centering
  \includegraphics[width=0.92\textwidth]{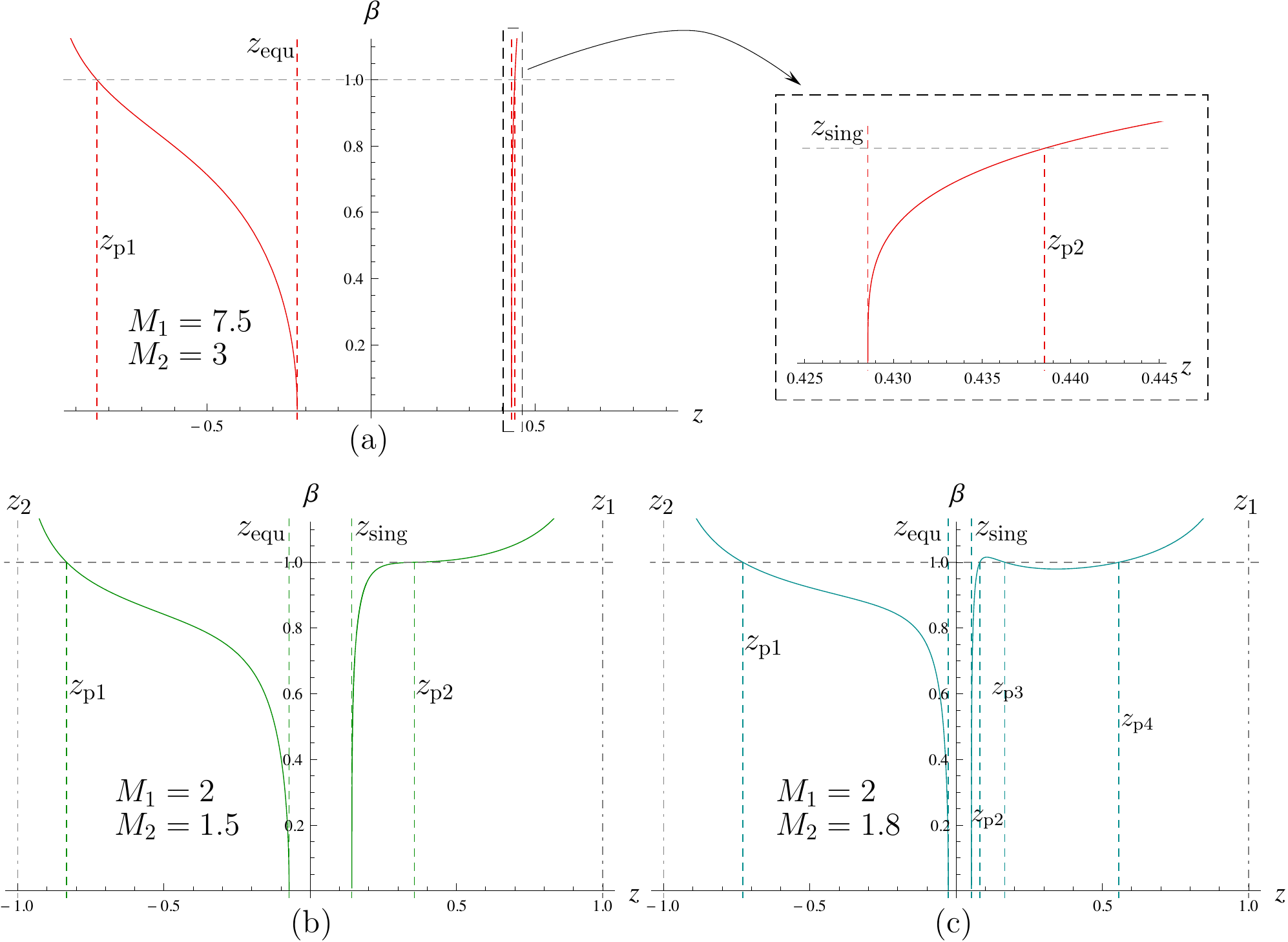}
  \caption{(color online) Local velocity $\beta(z)$ from Eq.~\eqref{equ:MP2MetricGeneralConditionForLocelVelocityOnCircularGeodesic} for a circular orbit at height $z$. Shown are plots for several combinations of the masses $M_1$ and $M_2$. The domain with existing photon orbits is bounded by the photon orbits discussed in Sec.~\ref{sec:lightlikeCircularOrbitsOutsideEquatorialPlane} and the characteristic heights $z_{\text{equ}}$ and $z_{\text{sing}}$.}  
  \label{pic:MP2MetricTimelikeOrbitsOutsideEquatorialPlaneForUnequalMasses}
\end{figure}
\twocolumngrid
\ \\
\ \\
\ \\

\subsection{Some explicit cases}   \label{sec:someConcreteCases}

\begin{figure}
  \centering
  \includegraphics[width=0.48\textwidth]{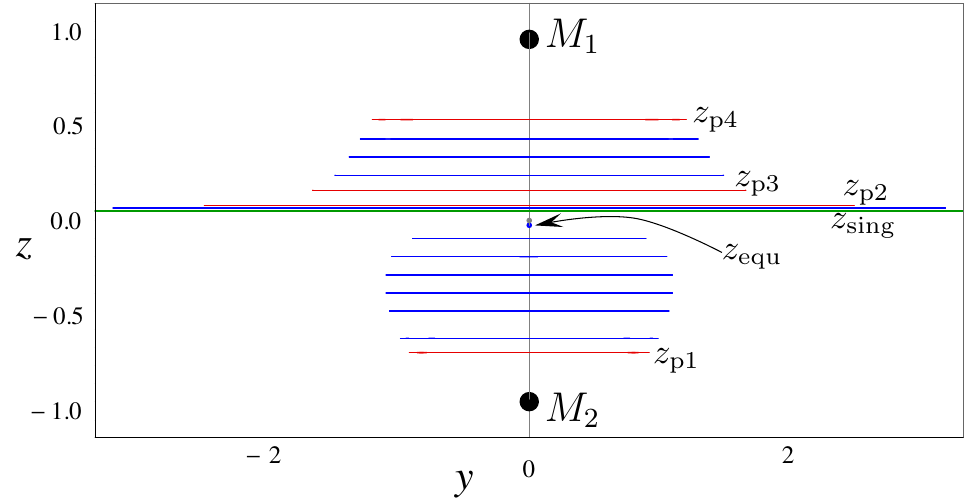}
  \caption{(color online) Edge-on view of circular orbits for $M_1=2$ and $M_2=1.8$. There are four photon orbits with heights ${z_{\text{p}1}\approx -0.72907}$, ${z_{\text{p}2}\approx 0.08139}$, ${z_{\text{p}3}\approx 0.16681}$, and ${z_{\text{p}4}\approx 0.55673}$, and radii ${\rho_{\text{p}1}\approx 1.00515}$, ${\rho_{\text{p}2}\approx 2.72863}$, ${\rho_{\text{p}3}\approx 1.81819}$, and ${\rho_{\text{p}4}\approx 1.31671}$. The plane with ${z=z_{\text{sing}}\approx 0.05263}$ marks the orbit with infinite radius and vanishing velocity. The point on the $z$ axis at ${z=z_{\text{equ}}\approx 0.02633}$ is the circular orbit degenerated to a point. In this case, there are three $z$ domains, where timelike circular orbits are possible inside.}
  \label{pic:MP2MetricCircularOrbitsOutsideEquatorialPlaneExample3}
\end{figure}

We conclude with the presentation of three examples that show the three-dimensional structure of lightlike and timelike circular orbits outside the equatorial plane of the extreme RN dihole metric. Figures~\ref{pic:MP2MetricCircularOrbitsOutsideEquatorialPlaneExample1}--\ref{pic:MP2MetricCircularOrbitsOutsideEquatorialPlaneExample3} were created using the GeodesicViewer by M{\"u}ller and Grave~\cite{MuellerGViewer2010}. This interactive visualization tool numerically integrates the geodesic equation
%\begin{equation}
%  \frac{d^2x^{\mu}}{d\lambda^2}+\Gamma^{\mu}_{\alpha\beta}\frac{dx^{\alpha}}{d\lambda}\frac{dx^{\beta}}{d\lambda} = 0
%\label{equ:geodesicEquation}
%\end{equation}
%
with initial conditions with respect to a local tetrad system, Eq.~\eqref{equ:MP2MetricCylindricalCoordinatesNaturalTetrad}. The heights $z_{\text{p}i}$ of the photon orbits follow from the numerical solutions of Eqs.~\eqref{equ:MP2MetricCircularOrbitsOutsideEquatorialPlaneEqualMassesUglyCondition} or~\eqref{equ:MP2MetricCircularOrbitsOutsideEquatorialPlaneUnequalMassesUglyCondition}, respectively, and the corresponding radii $\rho_{\text{p}i}$ from Eq.~\eqref{equ:MP2MetricCircularOrbitsOutsideEquatorialPlaneRhoSquaredFromZ}. For the timelike orbits, a $z$ value from an allowed domain is given; the corresponding radius $\rho$ and velocity $\beta$ then follow from Eqs.~\eqref{equ:MP2MetricCircularOrbitsOutsideEquatorialPlaneRhoSquaredFromZ} and~\eqref{equ:MP2MetricGeneralConditionForLocelVelocityOnCircularGeodesic}. 
% \begin{figure}
%   \centering
%   %\includegraphics[width=0.48\textwidth]{exampleOrbits1}
%   \includegraphics[width=\standardSize\textwidth]{wuensch_figure27}
%   \caption{(color online) Edge-on view of circular orbits for $M_1=M_2=1.5$. The photon orbits are located at ${z_{\text{p}\pm}\approx\pm 0.80324}$ with $\rho_{\text{p}\pm}\approx 0.95502$. In the plane ${z=z_{\text{sing}}=z_{\text{equ}}=0}$, there are no photon orbits but an arbitrary number of timelike orbits (compare with Secs.~\ref{sec:lightlikeCircularOrbitsInEquatorialPlane} and~\ref{sec:timelikeCircularOrbitsInEquatorialPlane}).}
%   \label{pic:MP2MetricCircularOrbitsOutsideEquatorialPlaneExample1}
% \end{figure}
% %
\begin{figure} \setcounter{figure}{27}
  \centering
  \includegraphics[width=0.48\textwidth]{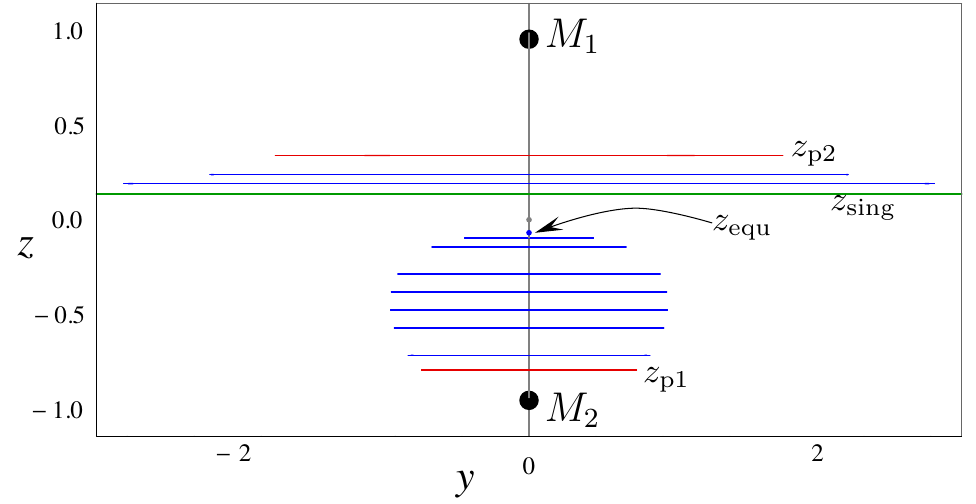}
  \caption{(color online) Edge-on view of circular orbits for $M_1=2$ and $M_2=1.5$. The photon orbits are located at $z_{\text{p}1}\approx 0.832490$ with $\rho_{\text{p}1}\approx 0.80124$ and $z_{\text{p}2}\approx 0.35591$ with $\rho_{\text{p}2}\approx 1.89345$. The plane with $z=z_{\text{sing}}\approx 0.14286$ marks the orbit with infinite radius and vanishing velocity. The point on the $z$ axis at $z=z_{\text{equ}}\approx -0.07180$ is the circular orbit degenerated to a point.}
  \label{pic:MP2MetricCircularOrbitsOutsideEquatorialPlaneExample2}
\end{figure}
%

%% ------------------------------------------------------------------------
%%         Conclusion
%% ------------------------------------------------------------------------
\section{Conclusion}   \label{sec:conclusion}

We have analyzed the existence and structure of circular orbits in the extreme Reissner-Nordstr{\o}m dihole metric for lightlike and timelike test particles and found big differences between the different cases. We have also shown that such orbits can only exist in heights $z$ between the two singularities. % and only in planes normal to and centers on their connecting line. Thus, they reflect the cylindrical symmetry of the spacetime. \\

In the classification of the circular orbits, at first, we had to distinguish between two cases. First, we have considered a test particle restricted to the equatorial plane of the dihole system with equal masses. In this case, the dynamics reduces to an effective two-body problem and the structure is similar to a single Reissner-Nordstr{\o}m naked singularity investigated by Pugliese \textit{et al.}~\cite{Ruffini2011a}. The number of photon orbits depends on the concrete mass value: There are no orbits for $M<\bar{M}$ and two for $M>\bar{M}$. For $M=\bar{M}$, we have exactly one photon orbit. Timelike circular geodesics are possible for all radii, except in the domain between the photon orbits, if such orbits exist. Furthermore, we have analyzed the stability behavior in the radial direction. In future work, this could be extended to a global stability analysis.

In the second case, we have focused on circular orbits outside the equatorial plane for arbitrary combinations of the two mass values. For both lightlike and timelike geodesics, their radii are determined by the same expression~\eqref{equ:MP2MetricCircularOrbitsOutsideEquatorialPlaneRhoSquaredFromZ} depending on the height of the orbit and on the mass ratio. We have found big differences for equal and unequal masses. In the case of equal masses, there are two nonequatorial photon orbits for $M<M^{'}$. Their heights delimit the range where timelike circular orbits are possible. For $M\geq M^{'}$, there are no photon orbits and thus no timelike circular geodesics. For unequal masses, in general, there is a forbidden domain between the characteristic heights $z_{\text{equ}}$ and $z_{\text{sing}}$. Depending on the combination of the two masses, we find between two and four photon orbits. Here, $z_{\text{equ}}$, $z_{\text{sing}}$, and the heights of the photon orbits are the boundaries of the ranges with possible timelike orbits. In the case of nonequatorial circular orbits, further stability analysis would also be of interest. The extreme Reissner-Nordstr{\o}m dihole metric is a static axisymmetric spacetime, and thus the paper of Bardeen~\cite{Bardeen1970Stability} could be helpful in this context.

In this article, we have studied neutral particles. Because of the charges of the two centers of the dihole system, trajectories of charged test particles would give an even deeper insight into the structure of this spacetime. Moreover, the Majumdar-Papapetrou spacetimes offer the possibility to study geodesics in more complex multi black hole configurations. For such discussions, the results of this work could be very helpful as reference values. Even though such multi black hole spacetimes are unlikely to be realized in nature, they could demonstrate the diversity of particle orbits in general relativity which we will consider in future work.

%Furthermore, the results of this article can be used as starting point for studying more difficult systems. Because of the charge of the two black holes, the motion of a charged test particle would be of interest. Moreover, the Majumdar-Papapetrou spacetimes give us the opportunity to add any number of extreme black holes and hence to study analytic multi black hole solutions of the Einstein equations. For such discussions, the results and characteristic sizes calculated in this article, for example the characteristic masses or the important role of the heights $z_{\text{equ}}$ and $z_{\text{sing}}$, could be helpful as references.

%% ------------------------------------------------------------------------
%%                               acknowledgments
%% ------------------------------------------------------------------------
%\begin{acknowledgments}
%The authors would like to thank Sebastian Boblest$^1$ for carefully reading this article.
%\end{acknowledgments}

%% ------------------------------------------------------------------------
%%                                  appendix
%% ------------------------------------------------------------------------
\appendix

\section{Timelike circular worldline in the extreme Reissner-Nordstr{\o}m dihole metric}\label{app:generalCircularWorldlineInTheMP2Metric}

The four-velocity $\bm{u}$ of a timelike particle can be given with respect to a local tetrad [see Eq.~\eqref{equ:MP2MetricCylindricalCoordinatesInitialDirectionRespectivelyNaturalTetrad}]. We consider a circular worldline ($\xi=\pi/2$) with no component in the $\bm{e}_z$ direction ($\chi=\pi/2$). Therefore, the components $u^{\rho}$ and $u^{z}$ have to vanish. The four-velocity then reads
\begin{equation}
  \bm{u} = c\gamma \left(\bm{e}_{(t)} + \beta \bm{e}_{(\varphi)} \right) = \underbrace{\gamma U(\rho,z)}_{=\ u^t} \partial_t + \underbrace{\frac{\beta\gamma}{\rho U(\rho,z)}}_{=\ u^{\varphi}} \partial_{\varphi}
\label{equ:MP2MetricFourVelocityOnCircularWorldline}
\end{equation}
with $U(\rho,z)$ from Eq.~\eqref{equ:MP2MetricUFunktionCylindricalCoordinates}. The four-velocity $\bm{u}$ can be formally integrated to the circular worldline $x^{\mu}(\lambda)$
%\begin{equation}
%   x^{\mu}(\lambda) = \left( t(\lambda)=\gamma U(\rho,z)\lambda,\ \rho=\text{const},\ \varphi(\lambda)=\frac{\beta\gamma}{\rho U(\rho,z)}\lambda,\ z=\text{const}  \right),
%\label{equ:MP2MetricCircularWorldline}
%\end{equation}
%
parametrized by the affine parameter $\lambda$. The four-velocity $\bm{u}$ is independent of $\lambda$ because $\rho$ and $z$ are constant on the circular worldline. The four-acceleration can be calculated according to
\begin{equation}
   a^{\mu} = \frac{du^{\mu}}{d\lambda} + \Gamma^{\mu}_{\alpha \beta}u^{\alpha}u^{\beta}
\end{equation}
with the Christoffel symbols of the second kind $\Gamma^{\mu}_{\alpha \beta}$. This calculation yields $a^t=a^{\varphi}=0$ and
\begin{subequations}
\begin{align}
   a^{\rho} &= -\frac{\gamma^2}{\rho U^3} \left[\rho(1+\beta^2)\frac{\partial U}{\partial\rho}+\beta^2 U\right] , \label{equ:MP2MetricCircularWorldlineFourAccelerationRho}\\
   a^z &= \frac{1}{U^3}(1-2\gamma^2)\frac{\partial U}{\partial z}. \label{equ:MP2MetricCircularWorldlineFourAccelerationZ}
\end{align}
\end{subequations}
As we are interested in geodesic worldlines, the four-acceleration has to vanish. From $a^z=0$, we obtain $\partial U/\partial z=0$, which leads to the general condition for circular orbits, Eq.~\eqref{equ:MP2MetricGeneralConditionForCircularOrbits}. Equation~\eqref{equ:MP2MetricCircularWorldlineFourAccelerationRho} can be solved to give
\begin{equation}
   \beta = \sqrt{\frac{U}{U+\rho \frac{\partial U}{\partial\rho}}-1} 
\label{equ:AppendixMP2MetricGeneralConditionForLocelVelocityOnCircularGeodesic}
\end{equation}
which is an expression for the local velocity $\beta$ to remain on a circular orbit with radius $\rho$. With $U(\rho)$ from Eq.~\eqref{equ:MP2MetricUFunktionCylindricalCoordinatesEquatorialPlane}, we also obtain the formula for the local velocity in the equatorial plane from Eq.~\eqref{equ:MP2MetricTimelikeParticleInEquatorialPlaneLocalVelocityOnCircularOrbit}.

\section{Classical analogon for a massive test particle}\label{app:classicalAnalogon}

In this section, we give a brief overview of the classical analogon to the motion of a neutral and massive test particle in the extreme RN dihole spacetime.
\begin{figure}
  \centering
  \includegraphics[width=0.35\textwidth]{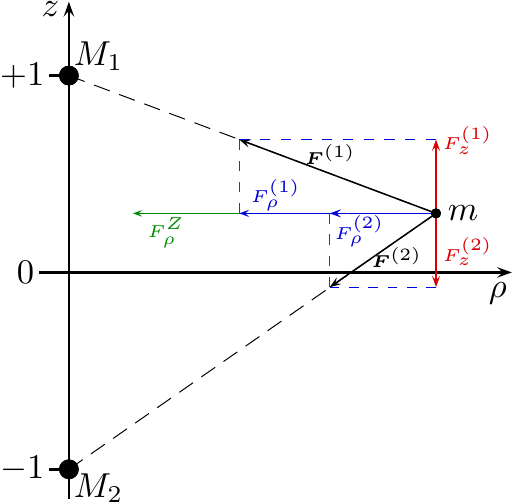} %0.34
  \caption{(color online) A neutral test particle with mass $m$ is attracted by the forces $\bm{F}^{(1)}$ and $\bm{F}^{(2)}$ of the two fixed masses $M_1$ and $M_2$. For describing a circular orbit, the $z$ components of the attracting forces have to compensate each other and the $\rho$ components have to act as centripetal force $\bm{F}^{Z}=F^{Z}_{\rho} \bm{e}_{\rho}$.}
  \label{pic:appendixMP2MetricClassicalAnalogon}
\end{figure}
In Fig.~\ref{pic:appendixMP2MetricClassicalAnalogon}, there is a sketch of this classical system. The masses $M_1$ and $M_2$ have charges $Q_1$ and $Q_2$ of the same sign. Both have the ratio $Q_i/M_i=1$ (Gaussian and geometric units); thus, the gravitational attraction is exactly compensated by the electrostatic repulsion. The mass $m$ of the test particle is negligible compared to the other masses. Therefore, $M_1$ and $M_2$ can be considered as fixed.

The test particle is attracted by the forces
\begin{subequations}
\begin{align}
   \bm{F}^{(1)} &= \frac{- \ m \ M_1 \ \rho}{\left[\rho^2+(z-1)^2\right]^{\frac{3}{2}}} \bm{e}_{\rho} - \frac{m \ M_1 \ (z-1)}{\left[\rho^2+(z-1)^2\right]^{\frac{3}{2}}} \bm{e}_{z} , \label{equ:classicalAnalogonForceOne} \\
   \bm{F}^{(2)} &= \frac{- \ m \ M_2 \ \rho}{\left[\rho^2+(z+1)^2\right]^{\frac{3}{2}}} \bm{e}_{\rho} - \frac{m \ M_2 \ (z+1)}{\left[\rho^2+(z+1)^2\right]^{\frac{3}{2}}} \bm{e}_{z} , \label{equ:classicalAnalogonForceTwo}
\end{align}
\end{subequations}
with the base vectors $\{\bm{e}_{\rho},\bm{e}_{\varphi},\bm{e}_{z}\}$ in cylindrical coordinates. Note that, in this section, bold letters are three-dimensional vectors. To describe a circular orbit, the $z$ components of the attracting forces must compensate each other, which results in the expression
\begin{align}
   - \frac{M_1 \ (z-1)}{\left[\rho^2+(z-1)^2\right]^{\frac{3}{2}}} = \frac{M_2 \ (z+1)}{\left[\rho^2+(z+1)^2\right]^{\frac{3}{2}}}. 
\label{equ:classicalAnalogonCompensatingZComponents}
\end{align}
Equation~\eqref{equ:classicalAnalogonCompensatingZComponents} is equivalent to the condition $\partial U/\partial z=0$ with $U(\rho,z)$ from Eq.~\eqref{equ:MP2MetricUFunktionCylindricalCoordinates}. This condition is already known from Sec.~\ref{sec:circularOrbitsInEquatorialPlane} and can be rearranged to Eq.~\eqref{equ:MP2MetricGeneralConditionForCircularOrbits}. Thus, in this context, the classical system has a similar behavior as in the general relativistic case: Circular orbits exist only in the range $z\in (-1;+1)$ and, for $z\neq 0$, we find an explicit connection between the radii $\rho$ and the heights $z$ of these orbits [see Eq.~\eqref{equ:MP2MetricCircularOrbitsOutsideEquatorialPlaneRhoSquaredFromZ} and Fig.~\ref{pic:rhoFromZForSeveralQ}]. For $z=0$ and $M_1=M_2$, the dynamics reduces to an effective two-body problem in the equatorial plane.
\begin{figure*}
  \centering
  \includegraphics[width=0.725\textwidth]{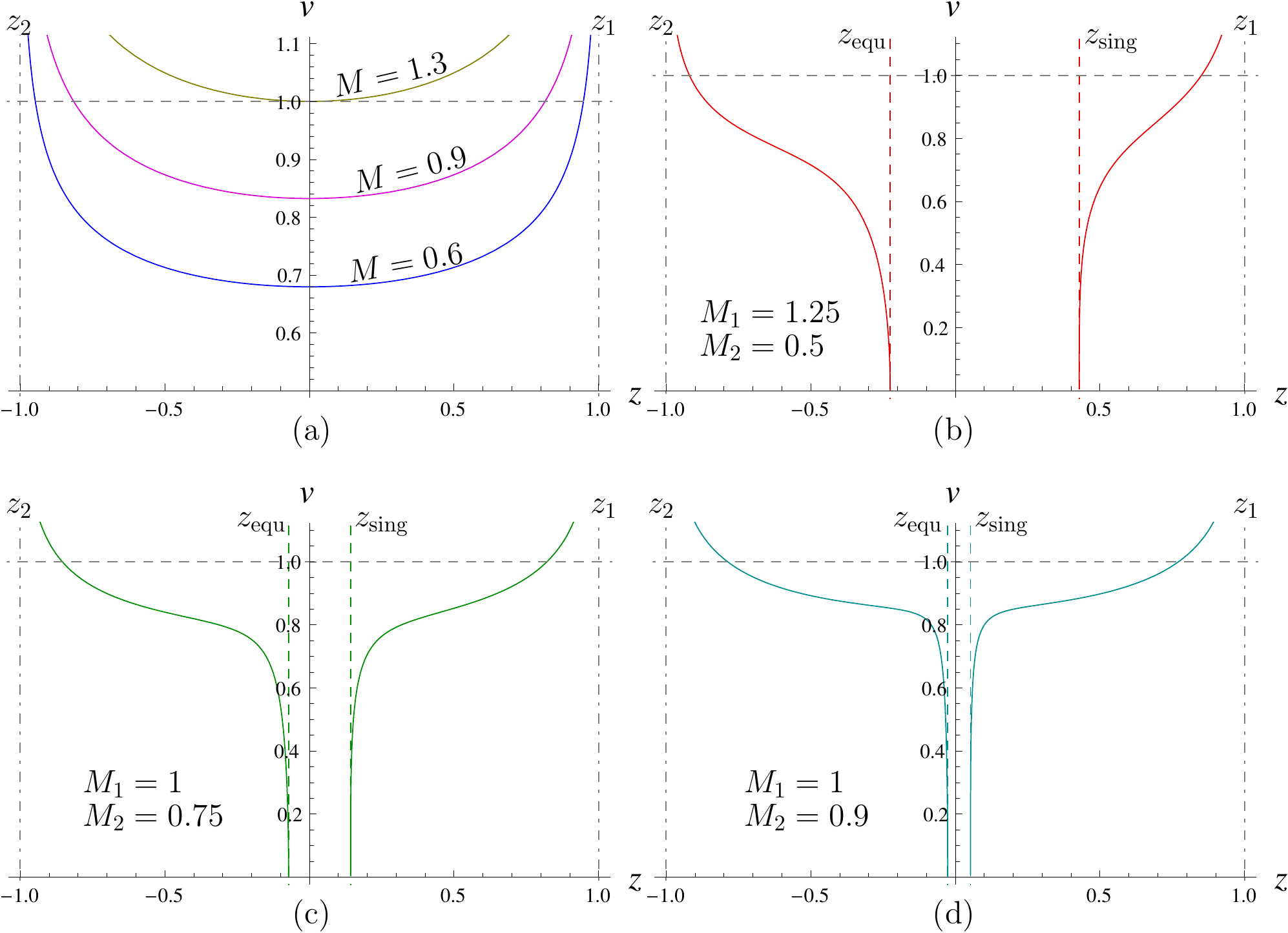} %0.5
  \caption{(color online) Classical velocity $v(z)$ from Eq.~\eqref{equ:classicalAnalogonVelocityOnCircularOrbit} for a circular orbit at height $z$ for (a) equal masses $M:=M_1=M_2$ and (b)--(d) several combinations of $M_1$ and $M_2$ with $M_1\neq M_2$. As in the relativistic case, the domain between the characteristic heights $z_{\text{equ}}$ and $z_{\text{sing}}$ is forbidden. In (b)--(d), the ratios $M_1/M_2$ correspond to those from Fig.~\ref{pic:MP2MetricTimelikeOrbitsOutsideEquatorialPlaneForUnequalMasses}. The speed of light ($v=1$) is also shown, but represents, in this classical case, no upper boundary of the velocity.}
  \label{pic:classicalAnalogonVelocityForCircularOrbits}
\end{figure*}
Furthermore, to describe a circular orbit, the sum of the $\rho$ components of the attracting forces has to be the centripetal force (see Fig.~\ref{pic:appendixMP2MetricClassicalAnalogon}). The condition $F^{(1)}_{\rho}+F^{(2)}_{\rho}=-mv^2/\rho$, with the velocity $v$ on the circular orbit, leads to
\begin{align}
   v = \sqrt{-\rho \frac{\partial U}{\partial \rho}} . 
\label{equ:classicalAnalogonVelocityOnCircularOrbit}
\end{align}
In this equation, $\rho(z)$ and $U(\rho(z),z)$ from Eqs.~\eqref{equ:MP2MetricCircularOrbitsOutsideEquatorialPlaneRhoSquaredFromZ} and~\eqref{equ:MP2MetricUFunktionCylindricalCoordinates} have to be used. Figure~\ref{pic:classicalAnalogonVelocityForCircularOrbits} shows the velocity $v(z)$ on a circular orbit at height $z\neq 0$ for several combinations of $M_1$ and $M_2$. These functions show a behavior qualitatively similar to the relativistic ones of Figs.~\ref{pic:MP2MetricTimelikeOrbitsOutsideEquatorialPlaneForEqualMasses} and~\ref{pic:MP2MetricTimelikeOrbitsOutsideEquatorialPlaneForUnequalMasses}, but with the important difference that the speed of light, in this classical system, is not an upper boundary of the velocity. Thus, in the case of equal masses, there are circular orbits for all $M$ and, in the case of unequal masses, there are two areas with circular orbits for all $M_1$ and $M_2$.

% -----------------------------------------------------------------
%                           thebibliography
% -----------------------------------------------------------------
%\nocite{chandrasekhar1989a,cunningham1973}
\section*{References}
\bibliographystyle{apsrev4-1}

%\bibliography{lit_ern.bib}

\end{document}